\documentclass{jfm}
\usepackage{graphicx}
\usepackage{amsmath}
\usepackage{xcolor}
\usepackage{ulem}

\def \deg {^\mathrm{o}}
\def \be{\begin{equation}}
\def \ee{\end{equation}}
\def \ba{\begin{eqnarray}}
\def \ea{\end{eqnarray}}
\def \bse{\begin{subequations}}
\def \ese{\end{subequations}}
\def \pd{\partial}
\def \bnabla{\boldsymbol{\nabla}}
\def \bs{\boldsymbol}
\def \mc{\mathcal} 

\def \wt{\widetilde}
\def \ovl{\overline}
\def \bc{\begin{center}}
\def \ec{\end{center}}

\shorttitle{Preferential orientation of floaters drifting in water waves}
\shortauthor{W. Herreman, B. Dhote, L. Danion, F. Moisy}

\title{Preferential orientation of small floaters drifting in water waves}

\author{Wietze Herreman\aff{1}
  \corresp{\email{wietze.herreman@universite-paris-saclay.fr}},
  Basile Dhote\aff{1},
  Lucile Danion\aff{1}, \\
  \and Fr\'ed\'eric Moisy\aff{1}}

\affiliation{\aff{1}Universit\'{e} Paris-Saclay, CNRS, FAST, 91405, Orsay, France.}

\begin{document}

\maketitle

\begin{abstract}
Elongated floaters drifting in propagating water waves slowly rotate towards a preferential orientation with respect to the direction of incidence. In this article, we study this phenomenon in the small-floater limit $k L_x <  1 $, with $k$ the wavenumber and $L_x$ the floater length. Experiments show that short and heavy floaters tend to align longitudinally, along the direction of wave propagation, whereas longer and lighter floaters align transversely, parallel to the wave crests and troughs. We show that this preferential orientation can be modeled  using an inviscid Froude-Krylov model, ignoring diffraction effects. Asymptotic theory, in the double limit of small wave slope and small floater, suggests that preferential orientation is essentially controlled by the non-dimensional number $F = k L_x^2 / \overline{h}$, with $\overline{h}$ the equilibrium submersion depth. Theory predicts the longitudinal-transverse transition {for homogeneous parallelepipeds} at the critical value $F_c = 60$, in fair agreement with the experiments that locate $F_c = 50 \pm 15$.  Using a simplified model for a thin floater, we elucidate the physical mechanisms that control the preferential orientation. 
The longitudinal equilibrium for $F<F_c$ originates from a slight asymmetry between the buoyancy torque induced by the wave crests, that favors the longitudinal orientation, and that induced by the wave troughs, that favors the transverse orientation. The transverse equilibrium for $F>F_c$ arises from the variation of the submersion depth along the long axis of the floaters, which significantly increases the torque in the trough positions, when the tips are more submersed. 
\end{abstract}


\section{Introduction}

The motion of a floating body in gravity waves is a classical problem in fluid mechanics 
with evident applications in the domain of naval engineering~\citep{faltinsen1993sea,newman2018marine,falnes_ocean_2020}. At first order in wave  magnitude,  waves cause harmonic oscillations of the floating body in all six degrees of freedom, {displacements} (heave, surge, sway) and {orientation angles} (pitch, roll, yaw). At second order,  waves also cause a mean drift force and yaw moment on the {body} that affect surge, sway and yaw angle on long time-scales.  For small isotropic floaters, this mean motion reduces to the classical Stokes drift in the direction of the wave propagation~\citep{stokes_theory_1847,van_den_bremer_stokes_2018,calvert_mechanism_2021}, a problem that received considerable interest for the modeling of pollutant transport in the oceans~\citep{suaria_2021, yang_2023, sutherland_fluid_2023}.  For larger floaters of arbitrary shape, such as ships and floating structures, an angular drift  can change the floater's orientation with respect to the wave incidence and this slow reorientation can in turn modify its linear drift. The combined linear and angular drifts are key features in sea keeping and maneuvering \citep{skejic2008unified}, and their modeling has been the subject of numerous works. The main methods to analyse ship motion are summarized in the books of  \cite{faltinsen1993sea} and \cite{newman2018marine}. 

\begin{figure}
\includegraphics[width=\textwidth]{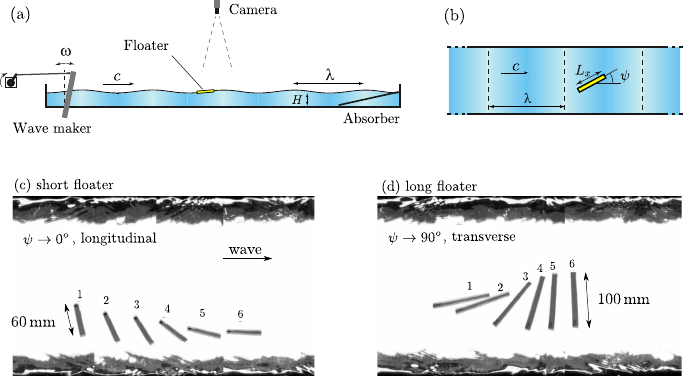} 
\caption{Experimental setup and chronophotographies. (a) Side view. Waves are generated in a water channel, of length 3~m and filled at height $H=10$~cm. (b) Top View. Floater of length $L_x$, making an angle $\psi$ with the direction of wave propagation. (c-d) Chronophotographies, obtained by superimposing images acquired every wave period, for a wave length $\lambda = 29$~cm and wave slope $\epsilon = ak = 0.16$. The short floater (length $L_x = 60$~mm) gradually aligns in the direction of wave propagation (c), while the long floater ($L_x = 100$~mm) aligns parallel to the wave crests (d).}
\label{fig:setup}
\end{figure}

In this paper, we are specifically interested in the slow, second order yaw motion of small elongated floaters drifting in gravity waves and how it creates a preferential state of orientation with respect to the incident waves. To illustrate this phenomenon, we show in figure~\ref{fig:setup} two chronophotographs from our laboratory experiments. Small homogeneous parallelepipeds of centimeter scale are left adrift in a propagating wave in a 3-m long water tank, and pictures synchronized with the wave period are taken from above (details are given in section 2). In just a few wave periods, we observe that  shorter floaters align with the direction of wave propagation, whereas longer floaters clearly prefer to align parallel to the wave crests. 
The main objective of this article to clarify what physically governs this preferential orientation.

In naval engineering contexts, the slow angular motion of elongated floaters is well known and sometimes referred to as low-frequency yawing. More than a century ago,  \cite{suyehiro_yawing_1921} reported how small boat models rotate towards a preferential state of orientation  and proposed a mechanism based on gyrostatic moments. Newman disagreed with this solid-mechanical explanation and proposed in 1967 his famous article that rationalised how ocean waves cause a mean drift force and moment on floating structures~\citep{newman_drift_1967}.  Starting from a global momentum and torque balance and using Green function theory, Newman expresses the second order drift force and yaw moment in terms of Kochin functions. These functions relate to the {\it far-field} limit of the hydrodynamic potential of the wave and and hence to how the wave is diffracted or modified by the moving floater.  Using slender body theory, Newman was able to calculate these Kochin functions and this lead him to propose an explicit formula for the mean yaw moment acting on slender floaters.  This theoretical mean yaw  moment compared reasonably well with the few experimental data-points of \cite{spens1962measurements} (reproduced in \cite{newman_drift_1967}). However, it also seems that some relevant physical effects are absent in this early theory: according to Newman's formula, slender floaters that are shorter than the wave-length should always be stable in {\it beam seas}, i.e., they should align their long axis parallel to the wave crests, in contradiction with Figure \ref{fig:setup}.

The far-field approach of Newman triggered many subsequent theoretical works on the topic of mean drift force and moment. \cite{salvesen1974second} is a simplification of Newman's model in which the floater is considered as a weak scatterer of the incoming wave.  In \cite{kashiwagi1992added} and \cite{kashiwagi1993study}, Parseval's theorem is used to evaluate the mean drift force and yaw moment, instead of the method of stationary phase  used by Newman. An entirely different, {\it near-field approach}, of direct pressure integration was proposed by \cite{faltinsen1980prediction}. Different theoretical methods are compared in \cite{skejic2008unified}, and show consistent results. \cite{chen2007middle} proposed a third method, the so-called {\it middle field formulation}, to compute mean wave loads on structures. A boundary element method is used to calculate the hydrodynamic potential and Kochin functions. This method is also implemented in the software pack Hydrostar~\citep{veritas2016hydrostar} that is specifically designed for naval engineering applications. Experiments on the specific subject of slow yawing are not so common.  In \cite{boulluec_steady_2008} it is briefly mentioned that elongated, container-like floaters can drift either in longitudinal positions (head-seas) or transverse position (beam-seas). Recently, \cite{yasukawa2019evaluations} compared the mean yaw moment obtained with a far-field theory to new experimental measurements on a particular ship model, and according to the authors the agreement is not so good. 

Overall, few experimental studies were specifically dedicated to the topic of slow yawing or preferential orientation and this was a first motivation to do this study. Secondly, we also want to better understand the physics that controls this preferential orientation. In this article, we propose a new experimental, numerical and theoretical study that is fully dedicated to the subject of preferential orientation of small floaters. When floaters are small with respect to the wavelength, diffraction is less important and preferential orientation likely simpler to understand. 

The article is structured as follows. In section~\ref{sec:expe}, we present a systematic series of experiments investigating the preferential orientation of small floaters of varying length and density. In section~\ref{sec:num}, we define an {idealised model for the  motion of small floaters in propagating gravity waves. We use the Froude-Krylov approximation that  ignores diffraction and we show that numerical solutions of this nonlinear model }reproduce well the observed state of preferential orientation in our experiments. In section~\ref{sec:theory}, we propose an asymptotic solution to this Froude-Krylov model in the double limit of small wave {slope  $\epsilon = ka$ (with $k$ the wavenumber and $a$ the wave amplitude)} and small floater size. {For elongated floaters with height $L_z$, width $L_y$ and length $L_x$ ordered as $kL_z \ll kL_y \ll kL_x \ll 1$, this} yields the idealized evolution equation for the average yaw angle $\ovl{\psi}$,
\be
\ddot{\ovl{\psi}} \approx - \epsilon^2 \sin \ovl{\psi} \cos^3 \ovl{\psi} \left ( 1 -  \frac{F}{F_c}  \right ).
\label{eq:psidd}  
\ee
{This equation depends on the non-dimensional number $F$, defined as}
\be
F = \frac{k L_x^2}{\beta L_z}, 
\label{eq:deff}
\ee
Here $\beta$ is the floater-to-water density ratio and we recognise $\bar h = \beta L_z$ as the equilibrium submersion depth or {\it draft} of our parallelepiped floaters. The idealised, dissipation-less evolution equation \eqref{eq:psidd} certainly does not capture all the complexity of realistic slow yawing, but {it provides insight into} the preferential state of orientation of small floaters. It suggests that, for $F < F_c$, the longitudinal position $\ovl{\psi}=0^o$ is stable and hence preferred,  whereas for $F > F_c$, the transverse position $\ovl{\psi}=90^o$ is stable. {Our theory predicts a transition at $F_c=60$, in fair agreement with the experimental value $F_c \simeq 50 \pm 15$. Relaxing the assumption $kL_y \ll kL_x$,  we obtain a more general theory that suggests bistability in a narrow interval slightly below $F_c$, {which may explain} in part the experimental uncertainty.}

The technicity of this asymptotic theory obscures physical insight and therefore, we {propose  in section~\ref{sec:needle} a second derivation of} equation \eqref{eq:psidd}  using a shorter procedure. This shows better what physically controls the preferential orientation. {The longitudinal equilibrium for short floaters arises through a mechanism similar to that explaining the Stokes drift of a material point. The buoyancy torque induced by wave crests rotates the floater in the longitudinal orientation, while that induced by wave troughs rotates the floater in the transverse orientation. Because of the slightly stronger buoyancy force on the crests than on the troughs, this produces a mean, second order, torque that favors the longitudinal orientation. Such a phase correlation between oscillating buoyancy force and oscillating level arm is analogous to that of {the} classical Kapitza pendulum, a pendulum whose anchor point is rapidly vibrated~\citep{kapitza_1951,landau,butikov_2001}, which tends to align along the direction of vibration}.
The transverse state of orientation is on the other hand due to the fact that long floaters have a variable submersion along their length.  This variable submersion significantly enhances the instantaneous yaw moment in trough positions that always favors the transverse position. 
 In section \ref{sec:comparison}, we finally compare our analytical formula for the mean yaw moment to previous results obtained by \cite{newman_drift_1967} and \cite{chen2007middle}, and show good agreement in the small floater limit $kL_x < 1$.  We find that in Newman's model, only the part of the mean yaw moment favoring the transverse position is present and this explains why his model does not predict a transition in preferential orientation.

\section{Experiments}
\label{sec:expe}

A series of laboratory experiments with centimeter-scale floaters of varying size and density have been performed in a  water flume. The experimental setup, sketched in figure~\ref{fig:setup}(a-b), consists in a tank of length 3~m, width 0.38~m, filled with water at height $H=0.1$~m. Waves are generated by a wavemaker consisting in a paddle oscillating at frequency $\omega/2\pi$ between 1 and 4~Hz, and are absorbed at the other end of the channel by a sloping plate.

We determine the wave profile $\zeta(x,t)$ by imaging the instantaneous contact line through the lateral channel wall and using a line detection algorithm.  The wave profile is well described by $\zeta = a(t) \cos(kx + \varphi(t))$, from which we determine the wave number $k$ and the instantaneous amplitude $a(t)$ and phase $\varphi(t)$. Because of imperfect wave attenuation at the sloping plate, the wave contains a small steady component, resulting in temporal oscillations of the amplitude $a(t)$ of the order of 5\%. From this fit, we determine the mean wave amplitude, simply denoted $a$ in the following. The measured wave number matches the dispersion relation in the gravity regime,  $\omega^2 = gk \tanh(kH)$, with wave lengths $\lambda$ in the range $10 - 82$~cm. Experiments for wavelengths of the order of the channel width are discarded because resonant transverse sloshing modes are excited. We work in the weakly nonlinear wave regime, for wave slopes $\epsilon = ak$ between 0.02 and 0.23. 

\begin{table}
\begin{center}
\begin{tabular}{p{3.2cm}p{1.8cm}p{1.8cm}p{1.8cm}p{1.8cm}p{1.8cm}}
     Material & $\beta$ & $L_y$ (mm) & $L_z$ (mm) & $h$ (mm) & $\beta L_z$ (mm) \\ 
     \\
	 PU foam & $0.26 \pm 0.02$ & $10.4 \pm 0.2$ & $ 4.9 \pm 0.05$ & $2.4 \pm 0.2$ & $1.3 \pm 0.1$ \\ 
	 PVC foam   & $0.44 \pm 0.02$ & $10.4 \pm 0.2$ & $4.9 \pm 0.05$ & $2.8 \pm 0.2$ & $2.1 \pm 0.1$ \\ 
	 PVC foam + Rubber & $0.73 \pm 0.02$ & $20.5 \pm 0.2$ & $6.9 \pm 0.05$  & $5.3 \pm 0.2$ & $5.0 \pm 0.1$ \\ 
	 ABS        & $0.90 \pm 0.02$ & $10.2 \pm 0.2$ & $4.3 \pm 0.05$  & $4.1 \pm 0.2$ & $3.9 \pm 0.1$ \\ 
\end {tabular}
\end{center} 
\caption{Floater properties: density ratio $\beta = \rho_s / \rho$, width $L_y$, thickness $L_z$, measured immersion depth $h$, predicted immersion depth without capillary effects $\beta L_z$. The floater length $L_x$ (not shown) is between 20 and 140~mm for each material. PU: Polyurethane. PVC: Polyvinyl Chloride. ABS: Acrylonitrile Butadiene Styrene.}
\label{Tab:1}
\end{table}

The floaters are homogeneous rectangular parallelepipeds of varying length $L_x$, in the range $20 - 140$~mm, and fixed width $L_y$ and thickness $L_z$. They are cut from various materials of floater-to-water density ratio $\beta = \rho_s / \rho$ between 0.26 and 0.90, as summarized in Table~\ref{Tab:1}. Without capillary effects, the expected equilibrium immersion depth $\beta L_z$ is in the range $1.3-5$~mm, but because of the hydrophilic nature of the materials, capillary forces tend to sink the floater slightly more. The immersion depth $h$ for each floater has been measured from the projection of a narrow laser sheet at a shallow incidence angle of $10^\mathrm{o}$ onto the floater positioned at the surface of water. The horizontal displacement of the laser sheet's intersection with both the water surface and the floater's surface was measured with accuracy $\pm 0.2$~mm.  The measured immersion depth $h$, given in Table~\ref{Tab:1},
consistently exceeds $\beta L_z$ by approximately 1 mm, in correct agreement with the expected capillary correction $2 \ell_c^2 \cos \theta / L_y$, with $\ell_c  = \sqrt{\gamma/\rho g} \simeq 2.5$~mm the capillary length, $\gamma$ the surface tension, and $\theta$  the contact angle.

\begin{figure}
\includegraphics[width=\textwidth]{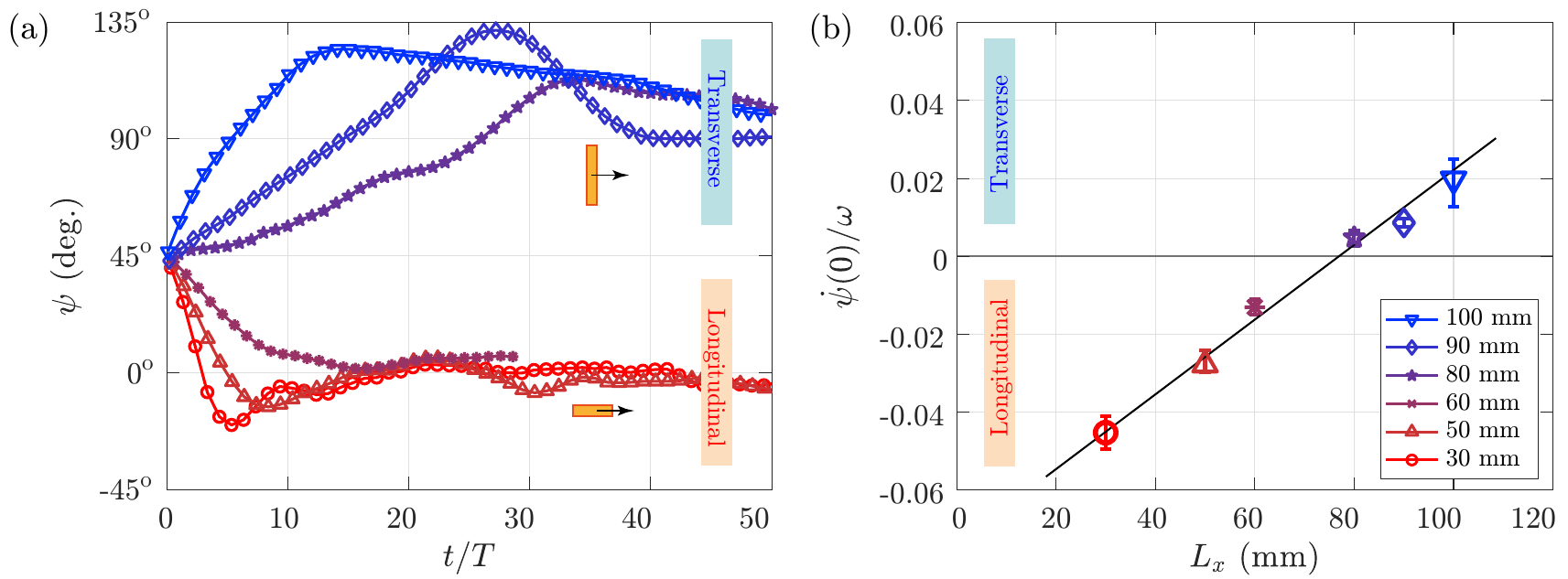} 
\caption{(a) Yaw angle $\psi$, {sampled at each wave period}, as a function of the time normalized by the wave period, for floaters of various lengths $L_x$. The time origin is chosen such that $\psi(0) = 45^\mathrm{o}$. Wave frequency $f = 2.3$~Hz, wavelength $\lambda = 0.29$~m, wave slope $ak = 0.16$, floater immersion depth $\bar h = 2.2$~mm. (b) Normalized angular velocity $\dot \psi/\omega$ {measured at $\psi = 45^\mathrm{o}$} as a function of $L_x$.}
\label{fig:ts}
\end{figure}

The equilibrium yaw angle of drifting floaters is determined as follows.  For each run, a floater is gently deposited at the surface of water at a distance $x_0 \simeq 0.8$~m from the wave maker, with an initial yaw angle $\psi_0$ of approximately $45\deg \pm 15\deg$ from the $x$ axis (controlling precisely $\psi_0$ is difficult because the phase of the wave is not known at the time the floater is released).  The floaters are imaged by a camera located above the wave tank. Using a tracking algorithm (library Tracker in Python), we measure the center of mass $x_c(t)$ and the yaw angle $\psi(t)$ of the floater on each frame. For each run, the floater is left adrift for about 1~m, and only trajectories staying approximately along the center line of the channel are retained to discard possible interaction with the side walls.

At first order, the floater motion is a combination of a back-and-forth oscillation of its center of mass, of amplitude given by the typical horizontal excursion of the fluid particle trajectories ($\Delta x = a / \tanh(kH)$ for waves in finite depth), and angular oscillations. Superimposed to these fast oscillations are a slow drift of the center of mass in the direction of the wave propagation (Stokes drift)  and a slow drift of the yaw angle, either towards the longitudinal ($\psi = 0 \deg$) or transverse ($\psi = 90 \deg$) orientation. To filter out the fast oscillations and focus on the slow dynamics of $\psi$, we synchronize the image acquisition with the wave maker oscillation, as illustrated in the chronophotographies of figure~\ref{fig:setup}(c-d). 

We first consider a set of floaters with a density ratio $\beta = \rho_s / \rho = 0.44$ cut from expanded PVC foam boards of thickness $L_z = 4.9$~mm, and investigate the influence of the floater length $L_x$,  for a fixed wavelength $\lambda = 290$~mm and wave amplitude $a=7.4$~mm (wave slope $\epsilon = ka = 0.16$). The time evolution of the yaw angle $\psi$, {sampled at the wave period,} is shown in figure~\ref{fig:ts}(a). Because of the uncertainty on the initial angle $\psi_0$, we shift the time origin so that $\psi$ is $45 \deg$ at $t=0$ for each run.  The curves clearly separate in two groups, with small floaters ($L_x \leq 60$~mm) tending to $\psi = 0 \deg$ (longitudinal) and long floaters ($L_x \geq 80$~mm) tending to $\psi = 90 \deg$ (transverse). We note that while short floaters precisely align in the longitudinal direction, with $\psi \simeq 0 \pm 5 \deg$ at large time, long floaters show larger variations around $\psi \simeq 90 \pm 30 \deg$. We provide in section \ref{sec:theory} a possible explanation for this behavior. 

\begin{figure}
\includegraphics[width=\textwidth]{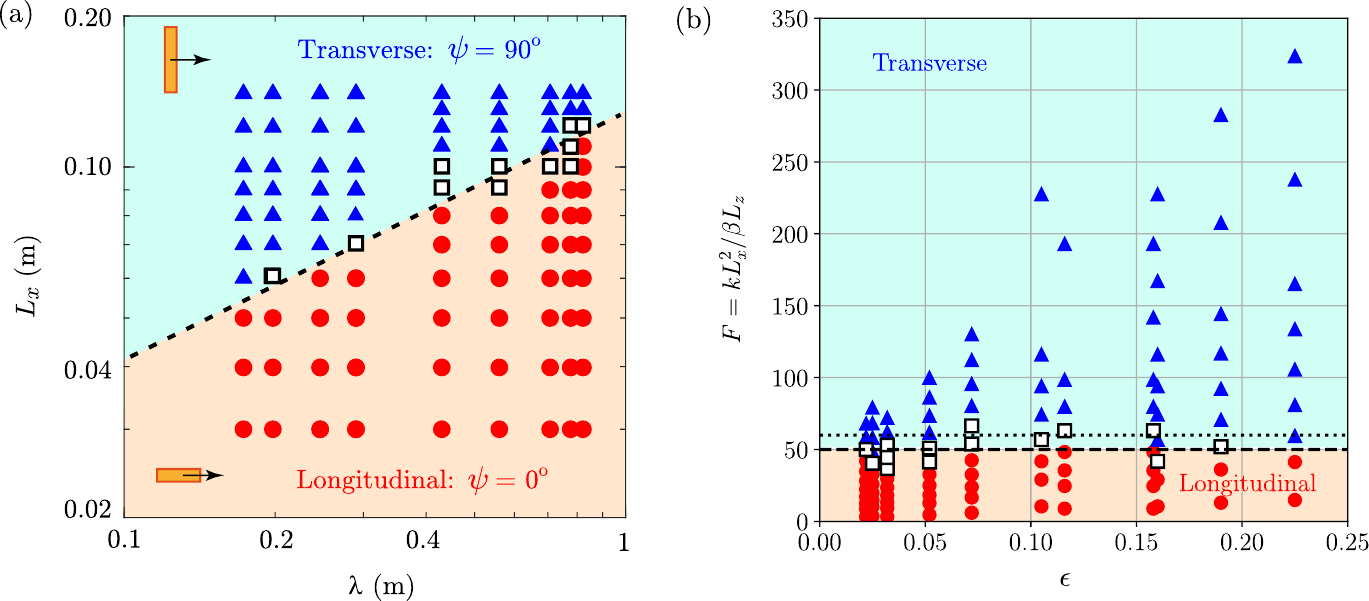} 
\caption{(a) Asymptotic floater orientation as a function of $L_x$ and wavelength $\lambda$, for floaters with density $\beta=0.44$, width $L_y=10$~mm and thickness $L_z = 5$~mm.  Red circles: longitudinal; Blue triangles: transverse. Black squares indicate indistinct states (erratic oscillations or non-reproducible experiments).  The separation line is $L_x = \sqrt{\ell \lambda}$, with $\ell \simeq 16 \pm 3$~mm. (b) Same data in the plan $(F, \epsilon)$, demonstrating  the independence of the orientation with the wave steepness $\epsilon$.  The long dashed line shows the experimental transition at $F_c \simeq 50$, and the short dashed line the theoretical prediction at $F_c = 60$.}
\label{fig:diag} 
\end{figure}

Figure~\ref{fig:ts}(a) shows that the reorientation dynamics is faster for very short or very long floaters: they reach their asymptotic orientation after approximately 5 wave periods only, while the convergence is much slower (at least 20 wave periods) for intermediate lengths. This convergence rate is illustrated in figure~\ref{fig:ts}(b) for various floater lengths, showing the angular velocity $\dot \psi$ measured at $t=0$ normalized by the wave frequency $\omega$  (this ratio measures the fraction of complete turn performed by the floater during one wave period). For this particular floater density and thickness, $\dot \psi / \omega$ crosses zero for $L_x \simeq 75$~mm, which defines the critical length $L_{xc}$ separating the longitudinal and transverse orientations. Because of the slow dynamics at the crossover, the orientation  is very sensitive to any experimental uncertainty for $L_x$ close to $L_{xc}$, such as the precise choice of the the initial angle $\psi_0$, inhomogeneities in the streaming flow, or small defects in the wetting line.

We have systematically determined the asymptotic yaw angle for various floater lengths $L_x$, wave lengths $\lambda$ and amplitudes $a$.  The preferential orientation of the floaters is first summarized in the plan $(L_x, \lambda)$ in figure~\ref{fig:diag}(a). When the asymptotic angle is $\psi \simeq 0 \pm 10 \deg$, floaters are labeled as ``longitudinal'' (red circles), and when $\psi \simeq 90 \pm 30 \deg$ they are labeled as ``transverse'' (blue triangles). Floaters with indistinct orientation are marked with black squares. This diagram shows a clear separation between the longitudinal and transverse orientations, with a  transition line well described by a square-root law, $L_{xc} \simeq \sqrt{\ell \lambda}$, with $\ell \simeq 16 \pm 3$~mm a fitting parameter. 

\begin{figure}
\begin{center}
\includegraphics[width=\textwidth]{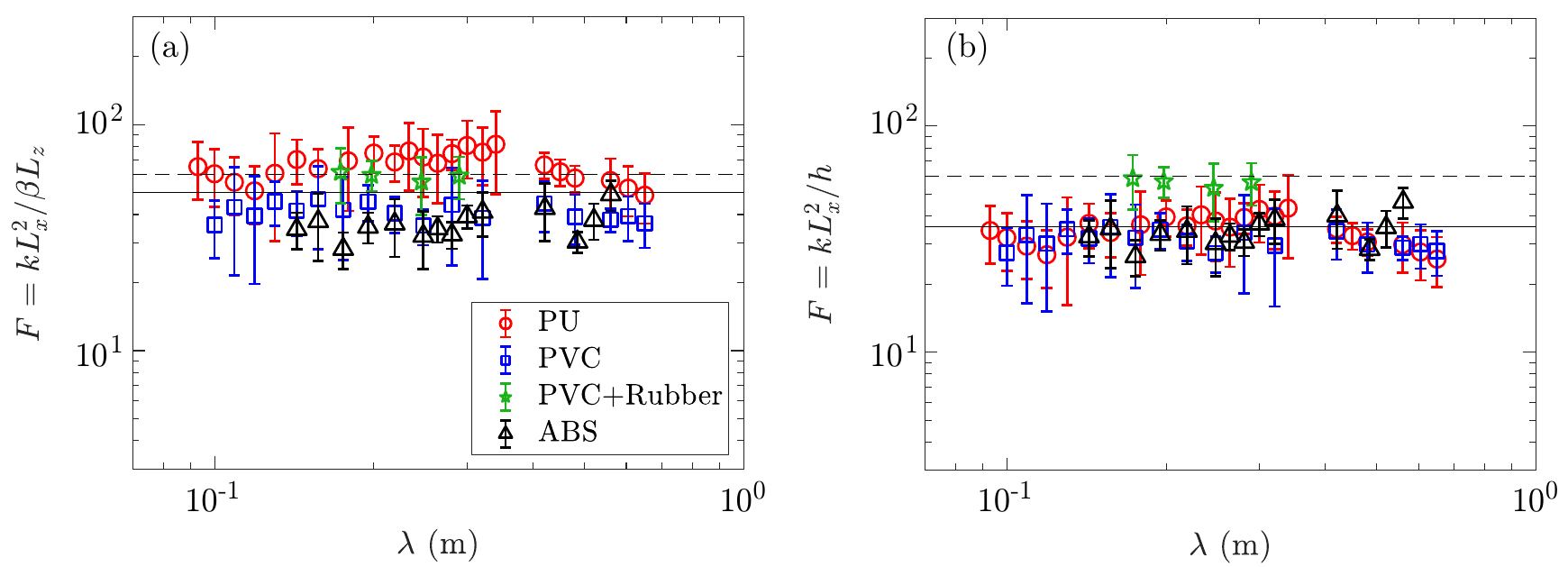}    
\end{center}
\caption{Critical values of $F = k L_{x}^2/\ovl{h}$ at the longitudinal-transverse transition as a function of the wavelength, for floaters of various densities (see Tab.~\ref{Tab:1}). (a), $F$ computed from the theoretical equilibrium immersion depth $\ovl{h}=\beta L_z$. The solid line shows the average value $F \simeq 50$. (b), $F$ computed from the measured immersion depth $\ovl{h}$ that is also affected by capillarity. The solid line shows the average value $F \simeq 35$. Dashed line: theoretical prediction $F_c = 60$.}
\label{fig:Lx2h_vs_lambda}
\end{figure}

This square-root law nicely conforms to the prediction of the asymptotic theory presented in the next section, which demonstrates that the preferential orientation is independent of the wave slope $\epsilon=ka$ and governed by the non-dimensional number $F=kL_x^2/\beta L_z$, with a longitudinal-transverse transition near $F_c = 60$. To check these predictions, we plot in figure~\ref{fig:diag}(b) the same data in the $(F,\epsilon)$ plane. We observe a clear separation between the longitudinal and transverse orientations, delimited by the line $F \simeq 50$, in fair agreement with the theory.

To further test the theory, we have determined the floater orientation for the 3 other materials listed in Tab.~\ref{Tab:1}, covering a range of densities $\beta$ from 0.26 to 0.90. For each wavelength $\lambda$, we determine the critical floater length $L_{xc}$ separating the longitudinal and transverse orientation, defined as the average between the largest longitudinal and smallest transverse floaters. Figure~\ref{fig:Lx2h_vs_lambda} summarizes the values of the critical parameter $F_c = k L_{xc}^2/h$ at the transition as a function of the wavelength, using either the predicted immersion depth without capillary effects $\beta L_z$ (Fig.~\ref{fig:Lx2h_vs_lambda}a), or the measured immersion depth $h$ (Fig.~\ref{fig:Lx2h_vs_lambda}b). The uncertainty on $L_{xc}$ is $\pm 5$~mm, leading to an uncertainty in $F_c$ of up to $\pm 20\%$ for each floater.  Both yield a constant value at the transition, $F_c\simeq 50 \pm 15$ in case (a) and $F_c \simeq 35 \pm 10$ in case (b), with a slightly reduced scatter in case (b).

{We conclude that the non-dimensional parameter $F$  successfully discriminates between the longitudinal and transverse orientations, in decent agreement with the theory, although the transition occurs at a value slightly smaller than the theoretical prediction $F_c = 60$. We introduce in the following section the simplifying assumptions of our theory, and discuss in the conclusion to what extent these assumptions could explain the discrepancy between experiments and theory.}

\section{Inviscid Froude Krylov model}
\label{sec:num}

The preferential state of orientation of small floaters can be explained using an idealised Froude-Krylov model that ignores diffraction. We define the equations of motion and solve them numerically to reproduce the main experimental observations on preferential orientation.  

\subsection{Simplifying assumptions}

In the model, we suppose that : 
\begin{enumerate}
\item  The incoming wave is a low amplitude, inviscid potential gravity wave in deep water --  In the experiments  the maximum wave slope is $\epsilon = ka = 0.23$; the deep-water assumption is not fully satisfied, with $\tanh(kH) \simeq 0.65$ for the largest wave-length. 

\item Viscosity is negligible --  The viscous stress on the floater is of the order $\eta a \omega / \delta_s$, where $\delta_s = (\nu/\omega)^{1/2}$ is the thickness of the Stokes boundary layer. This viscous stress is much smaller than the pressure variations in the wave, $ p \sim \rho a \omega^2 / k $, for $\nu k^2/\omega \ll  1$, which is well satisfied in the experiments: $\nu k^2/\omega  < 2 \times 10^{-4}$.

\item Capillarity is negligible -- The capillary length, $\ell_c = \sqrt{\rho g / \gamma} \simeq 2.5$~mm, is smaller than the characteristic floater size and wavelength in our experiments. We ignore {the change in equilibrium immersion depth $\overline{h}$ by the capillary forces}. 

\item  Steady streaming flows are negligible -- The steady streaming flow is a nonlinear, Eulerian mean flow correction of order $O(\epsilon^2)$ that comes along with the wave. This mean flow can affect reorientation, but only if it is inhomogenous at the scale of the floater. {Since} steady streaming flows typically vary spatially on the scale $k^{-1}$, we expect a weak effect of streaming on the floater orientation {for $kL \ll 1$}.

\item Wave scattering and emission are negligible -- Floaters are moving obstacles that scatter the incoming wave and that also emit waves. We ignore these flow modifications in our model,  an approximation that is known as the Froude-Krylov approximation. This is only reasonable when the floater is small with respect to the wavelength and when differential motion is weak: 
\be
\delta= k L  \ll 1 \quad , \quad  \frac{|| \bs{u} - \bs{v} ||}{|| \bs{u} ||} \ll 1,
\ee
with $\bs{u} $ is the fluid velocity and $\bs{v}$ the floater velocity. The Froude-Krylov approximation is not without physical consequences. By not considering diffraction, we filter out all radiative losses and added mass effects. One consequence is that free {``bobbing''} oscillations of the floater around its equilibrium position are not damped in our model.  Another consequence of the absence of dissipation is that, as for an undamped pendulum, {the equilibrium states must be} either unstable or {\it marginally stable}. {Our model therefore predicts oscillations around the stable fixed points, rather than a convergence towards them as in the experiments.}

\end{enumerate}

\begin{figure}
\begin{center}
\includegraphics[width=\textwidth]{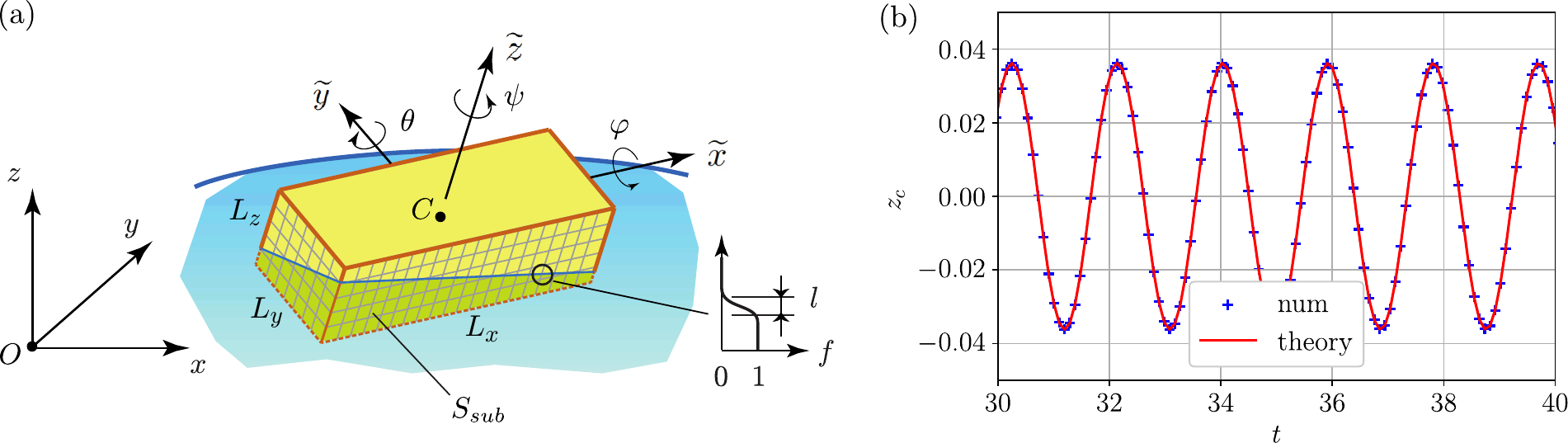} 
\end{center}
\caption{(a) Sketch of the floater and notations: laboratory frame $(O,x,y,z)$, moving floater frame $(C,\wt{x},\wt{y},\wt{z})$, and Euler angles $\theta$ (pitch), $\phi$ (rall) and $\psi$ (yaw). In the simulation, we numerically calculate the pressure force and moment on the submerged surface $S_{sub}$ using rectangular meshes on each face and a mask function $f$ that indicates whether the point on the face is submerged or not.  (b) Free vertical oscillations of $z_c$ at  the bobbing frequency $\sqrt{1/\beta \delta_z}$ in a numerical test-case without incoming wave 
\label{fig:mesh}
}
\end{figure}

\subsection{Incoming wave}

We define the incoming wave in the laboratory frame of reference $(O,x,y,z)$, with $(\bs{e}_x, \bs{e}_y,  \bs{e}_z)$ the basis vectors (figure~\ref{fig:mesh}). By convention, the origin $O$  is on the equilibrium fluid surface. The wave propagates along $x$ and is $y$-invariant. Up to first order in wave magnitude, the potential wave solution on infinitely deep water is
\bse  \label{h_ux_ux_p_def_dim_all}
\be \label{h_ux_ux_p_def_dim}
\zeta^{(1)}= a  \sin (kx- \omega t ) , \quad \left \{ 
\begin{array}{rcl} 
u_x^{(1)} &= &  a \omega e^{kz}   \sin (kx- \omega t)  \\
u_z^{(1)}  &= & - a \omega  e^{kz}  \cos (kx- \omega t)  \\
p^{(1)}&=& p_0 -  \rho g z +  \rho g a   e^{kz}  \sin (kx - \omega t),
\end{array} \right .
\ee
with $\zeta$ the surface elevation, $u_x,u_z,p$ the velocity components and pressure, $p_0$ the atmospheric pressure, and $\omega = \sqrt{gk}$. Since we are interested in second order effects in the floater motion, we also include second order corrections in the incoming wave. At second order in wave-magnitude, we have Stokes wave corrections of the surface and pressure but no extra flow,
\be
\zeta^{(2)} = - \frac{a^2 k}{2} {\cos (2 (kx- \omega t) )  } \ , \  \ p^{(2)}=  - \frac{\rho g a^2 k}{2} e^{2 k z} \ , \ u_x^{(2)}= u_z^{(2)}=0 .
\ee
\ese
The pressure is slightly reduced everywhere under the wave and the surface locally steepens at wave-crests and flattens at wave troughs.  

In the following, we non-dimensionalise space, time, velocity and pressure using the scales
\be
[ \bs{r} ]   = k^{-1} \quad , \quad [t] = (gk)^{-1/2} \quad , \quad  [\bs{u}] = g^{1/2} k^{-1/2}  \quad , \quad  [p] = \rho g k^{-1}  .
\ee
In this unit system, the wave slope $\epsilon = ka \ll 1$ is the only remaining parameter that defines the wave and equations~\eqref{h_ux_ux_p_def_dim_all} become
\bse  \label{h_ux_ux_p_def}
\ba
\zeta^{(1)} &=&  \epsilon \sin (x- t ) , \quad \left \{ 
\begin{array}{rcr} 
u_x^{(1)} &= &  \epsilon e^{z}   \sin (x- t)  \\
u_z^{(1)}  &= & - \epsilon e^{z}  \cos (x- t)\\
\quad  p^{(1)}&=& p_0 -  z +  \epsilon  e^{z}  \sin (x - t)
\end{array} \right .  \\
\zeta^{(2)} &=& - \frac{\epsilon^2}{2} {\cos (2 (x-  t) )  }    \ , \ p^{(2)}=  - \frac{\epsilon^2}{2} e^{2  z} \ , \ u_x^{(2)}= u_z^{(2)}=0.
\ea
\ese

\subsection{Equations of motion for the floater}

The floater is a rectangular parallelepiped with non-dimensional density $\beta$, non-dimensional length, width and height $( \delta_x , \delta_y, \delta_z) = (kL_x, kL_y,kL_z)$. We describe the instantaneous position of the center of mass $C$ of the floater relative to $O$. The position vector is decomposed in the {\it inertial, laboratory frame} :
\be
\bs{r}_c (t) =  \bs{OC}(t) = x_c( t) \bs{e}_x +y_c( t) \bs{e}_y +z_c( t) \bs{e}_z.
\ee
To describe the orientation of the floater, we introduce a second reference frame, the {\it non-inertial material frame } $(C,\wt{x},\wt{y},\wt{z})$, that is co-moving with the floater. By convention, the unit vectors $(\wt{\bs{e}}_x (t), \wt{\bs{e}}_y(t),  \wt{\bs{e}}_z(t))$ are aligned with the long, medium and short axes (figure~\ref{fig:mesh}). We introduce the three Euler angles, roll $\varphi (t)$, pitch $ \theta(t)$ and yaw $\psi(t)$, that connect the laboratory frame to the moving floater frame. In our angle convention (see supplementary material) this transform is  
\bse  \label{mante_to_lab}
\be  \label{mante_to_lab_base}
\left [ \begin{array}{c}
\bs{e}_x \\ \bs{e}_y  \\ \bs{e}_z  
 \end{array} \right ]  =    \underbrace{\left [ \begin{array}{ccc} 
 c_\psi c_\theta & \left ( c_\psi s_\theta s_\varphi -  s_\psi c_\varphi  \right ) & \left ( c_\psi s_\theta c_\varphi + s_\psi s_\varphi \right ) \\
s_\psi c_\theta & \left (s_\psi s_\theta s_\varphi + c_\psi c_\varphi  \right ) & \left (  s_\psi s_\theta c_\varphi - c_\psi s_\varphi \right )  \\
-s_\theta & c_\theta s_\varphi & c_\theta c_\varphi 
 \end{array} \right ]   }_{R^T } 
  \left [ \begin{array}{c}
\wt{\bs{e}}_x \\ \wt{\bs{e}}_y  \\ \wt{\bs{e}}_z    
 \end{array} \right ]  .
\ee
Here and further we denote in short $c_\psi = \cos \psi $, $s_\psi = \sin \psi $ and similar for the other angles.  This transform is easily inverted because $R$ is an orthogonal matrix, $R^{-1} = R^T$. Components of a vector $\bs{A}$ and the coordinates of the laboratory and floater frame are also connected by this matrix,
\be \label{mante_to_lab_coor}
\left [ \begin{array}{c}
A_x \\ A_y  \\ A_z   
 \end{array} \right ]  =   R^T  
  \left [ \begin{array}{c}
\wt{A}_x \\ \wt{A}_y  \\ \wt{A}_z  
 \end{array} \right ]   \quad, \quad  \left [ \begin{array}{c}
x  - x_c \\ y  -y_c \\ z  - z_c
 \end{array} \right ]  =  R^T\left [ \begin{array}{c}
\wt{x} \\ \wt{y} \\ \wt{z}   
 \end{array} \right ] .
\ee
\ese

In the absence of waves, $\epsilon=0$, the floater is in an equilibrium position, meaning perfectly leveled and with center $C$ that is $\delta_z/2$ above the bottom face submerged at depth $\overline{h} = \beta \delta_z$. Hence, we have
\be \label{eq:equilibrium}
\ovl{x}_{c} , \  \ovl{y}_{c} \ , \  \ovl{\psi} \quad \text{arbitrary} \quad, \quad  \ovl{z}_c = \left ( \frac{1}{2} - \beta \right ) \delta_z  \quad, \quad    \ovl{\theta} = \ovl{\varphi}  = 0 
\ee
at equilibrium. Here and further, we will use overbars to label quantities that do not vary on the short time-scale of the wave. 

In the presence of waves, the floater will be displaced and by definition, it moves as a solid. An arbitrary point $\bs{r}$ of the floater has velocity 
\be
\bs{v} =  \bs{v}_c (t)  +  \bs{\Omega} (t) \times ( \bs{r}  - \bs{r}_c  (t) ),
\ee
with $\bs{v}_c  (t)  = \dot{\bs{r}}_c (t)$ the  translation velocity of the center of mass and $ \bs{\Omega} (t)$ the instantaneous rotation velocity. 
We decompose $\bs{v}_c  (t) $ in the inertial, laboratory frame. The rotation vector  is decomposed in the non-inertial, floater frame: $ \bs{\Omega} (t) = \wt{\Omega}_x (t) \wt{\bs{e}}_x (t) + \wt{\Omega}_y(t)  \wt{\bs{e}}_y (t)  + \wt{\Omega}_z(t)  \wt{\bs{e}}_z (t) $. The components of this rotation vector are kinematically linked to the time-derivatives of the Euler angles (see supplementary material). We have 
\be \label{kinematics} 
\left [ \begin{array}{c}
\dot{x}_c \\  \dot{y}_c \\  \dot{z}_c 
 \end{array} \right ] = \left [ \begin{array}{c}
{v}_{c,x} \\  {v}_{c,y} \\ {v}_{c,z}
 \end{array} \right ] \ , \
\left [ \begin{array}{c}
\dot{\varphi} \\  \dot{\theta} \\  \dot{\psi } 
 \end{array} \right ]  = \left [ \begin{array}{ccc} 
 1 & \sin \varphi \tan \theta & \cos \varphi \tan \theta \\
  0 & \cos \varphi & - \sin \varphi  \\
  0 & \dfrac{\sin \varphi }{\cos \theta} & \dfrac{\cos \varphi}{\cos \theta}
  \end{array} \right ]  \left [ \begin{array}{c}
\wt{\Omega}_x \\ \wt{\Omega}_y \\  \wt{\Omega}_z
 \end{array} \right ] .
\ee 

From Newton's law and the angular momentum theorem, we have 
\be
m \dot{\bs{v}}_c =    \bs{F}   \ , \quad 
\frac{d}{dt} \left ( \bs{I} \cdot \bs{\Omega} \right ) =   \bs{K} ,
\ee
with  $m = \beta  \delta_x \delta_y \delta_z$ the non-dimensional mass of the floater and $ \bs{I} $ the non-dimensional inertia tensor. The principal moments of inertia of the rectangular parallelepiped are
\be
\wt{I}_{xx} = \frac{m (\delta_y^2 + \delta_z^2 )}{12}  , \quad \wt{I}_{yy} = \frac{m (\delta_z^2 + \delta_x^2 )}{12}  , \quad \wt{I}_{zz} = \frac{m (\delta_x^2 + \delta_y^2 )}{12} .
\ee
The floater is subject to its weight and to pressure forces and moments. In non-dimensional form, we have 
\be \label{FpKp}
\bs{F}=  \int_{S_{sub}} ( p  -p_0)  \,  d \bs{S}  -  m  \bs{e}_z \ , \quad 
\bs{K} =  \int_{S_{sub}} ( \bs{r} - \bs{r}_c) \times  (p  -p_0)  \,  d \bs{S} ,
\ee
with $d \bs{S} $ orientated towards the floater. In the Froude-Krylov approximation, we use the incoming wave pressure $p$ and surface elevation defined in equation \eqref{h_ux_ux_p_def}. Pressure and surface corrections due to scattered and emitted waves are entirely ignored. The submerged surface $S_{sub}$ corresponds to the part of the floater surface that has  $z< \zeta$. 
We decompose Newton's law  in the inertial laboratory frame and the angular momentum equation in the non-inertial floater frame. Using the fact that the material frame rotates along with the floater, i.e. that $\dot{\wt{\bs{e}}}_i (t) = \bs{\Omega}(t) \times \wt{\bs{e}}_i (t) $, we obtain
\be \label{dynamics}
\left [ \begin{array}{c}
m\dot{v}_{x,c} \\  m\dot{v}_{y,c} \\ m \dot{v}_{z,c} 
 \end{array} \right ] = \left [ \begin{array}{c}
F_{x} \\  F_{y}\\   F_{z}
 \end{array} \right ] \quad , \quad   \left [ \begin{array}{c}
\wt{I}_{xx} \dot{\wt{\Omega}}_x \\ \wt{I}_{yy}  \dot{\wt{\Omega}}_y \\  \wt{I}_{zz} \dot{\wt{\Omega}}_z
 \end{array} \right ] =  \left [ \begin{array}{c}
 {\wt{K}}_{x}  +  (  \wt{I}_{yy} -  \wt{I}_{zz} ) \wt{\Omega}_y {\wt{\Omega}}_z \\
{\wt{K}}_{y}  +   (  \wt{I}_{zz} -  \wt{I}_{xx} ) \wt{\Omega}_z {\wt{\Omega}}_x \\
{\wt{K}}_{z}  + (  \wt{I}_{xx} -  \wt{I}_{yy} ) \wt{\Omega}_x {\wt{\Omega}}_y 
  \end{array} \right ] .
\ee 
Equations  \eqref{dynamics} combined with the kinematic relations \eqref{kinematics} define a first order system of 12 differential equations. If we provide an initial state for the floater position and velocity, we can numerically integrate this system forward in time.

\subsection{Numerical simulations}

In our numerical code, we use the standard Runge-Kutta 4th order explicit numerical scheme. 
The numerical calculation of the surface integrals \eqref{FpKp}  that define the instantaneous $\bs{F}$ and $\bs{K}$ is non-trivial because the submerged surface $S_{sub}$ varies in time. We use the following procedure to compute these integrals (see sketch of figure \ref{fig:mesh}(a)).
On each of the six faces of the floater surface, we define two-dimensional rectangular meshes that contain $\wt{x},\wt{y},\wt{z}$ coordinates on those faces,  with typically $100 \times 100$ points. To evaluate the force components ${F}_{x}, {F}_{y},{F}_{z}$ at a given time $t$, we use a loop that visits all six faces. On each face, we first calculate the lab-frame $({x},{y},{z})$ coordinates of the points on that face, using the coordinate transform \eqref{mante_to_lab} and the present position $x_c(t) , y_c(t) , z_c(t) , \varphi (t), \theta (t), \psi (t)$.  With these lab-frame coordinates, we can evaluate the pressure $(p-p_0)$ on that face using equation \eqref{h_ux_ux_p_def}, {using either a first order description $p =p^{(1)}$ or a second order description, $p =p^{(1)}+ p^{(2)}$.}  To handle the fact that the faces can be totally, partially or not submerged, we also calculate an
indicator function  $f= (1+\tanh((\zeta(x,t)-z)/l)/2 $, with $l$  a length over which the interface is smoothed and $\zeta$ as in equation \eqref{h_ux_ux_p_def}, {using again either a first order description $\zeta =\zeta^{(1)}$ or a second order description, $\zeta=\zeta^{(1)}+ \zeta^{(2)}$.} This function $f$, equal to $1$ in the liquid and $0$ in the air, is a smooth numerical approximation of the Heaviside function. We then take the product $(p-p_0) f$ that is only non-zero on the submerged points of that face. Using a two-dimensional quadrature rule and the numerical values of $(p-p_0) f$ on the face, we can then compute the surface integral on that face. Each face gives a local contribution to the force in the direction of the local inward normal $d \bs{S} $, so after having visited all 6 faces, we obtain the components $\wt{F}_{x}, \wt{F}_{y},\wt{F}_{z}$ in the floater frame. Using the transform \eqref{mante_to_lab} we  obtain the force components ${F}_{x}, {F}_{y},{F}_{z}$ in the lab frame.  The moments $\wt{K}_{x}, \wt{K}_{y},\wt{K}_{z}$  are calculated similarly. 

We have done several static and dynamical tests in the absence of waves ($\epsilon=0$). In the static tests, we validated the calculation of  $\bs{F}$ and $\bs{K}$ on floaters that were submerged and rotated to positions for which we could easily compute the force and moment analytically.  In the dynamical tests, we used the code to reproduce free oscillations of the floaters. When we release the floater slightly off its equilibrium position \eqref{eq:equilibrium}, we expect free ``bobbing'' oscillations in the vertical $z_c$ or angular $\theta, \varphi$ coordinates, of non-dimensional frequencies \citep{falnes_ocean_2020}
\be \label{omdef}
\omega_z =\sqrt{\frac{1}{\beta \delta_z}}, \quad 
\omega_\theta = \sqrt{\frac{1}{\beta \delta_z} \left ( \frac{\delta_x^2 + 6 \beta (\beta-1) \delta_z^2 }{\delta_x^2 + \delta_z^2}   \right )},\quad  \omega_\varphi = \sqrt{\frac{1}{\beta \delta_z} \left ( \frac{\delta_y^2 + 6 \beta (\beta-1) \delta_z^2 }{\delta_y^2 + \delta_z^2}   \right )}.
\quad 
\ee
Figure \ref{fig:mesh}(b) shows an example of timeseries for free vertical oscillations in $z_c$, for a floater with $\beta=0.5$ released slightly above its equilibrium position $\ovl{z}_c=0$. As illustrated in the figure, the oscillatory motion $z_c = A \cos (\omega_z t)$  is accurately reproduced by our code.

The free oscillations in  $z_c, \theta, \varphi$ are  useful to test the code but do not reflect the behavior of the floaters in our experiment. Restricting to floaters much smaller than the wavelength, $\delta \ll 1$, implies that the bobbing frequencies \eqref{omdef} are much larger than the incoming wave frequency ($\omega=1$ in non-dimensional units). Accordingly, such fast bobbing oscillations are never resonant, and are expected to be rapidly damped in the experiments, either by viscous friction or radiation loss. Since no damping is included in our model, we need to minimize these parasitic bobbing excitations. We use for this the following strategy: At time $t=0$, we place the floater at its equilibrium position \eqref{eq:equilibrium}, and gradually ramp up the wave amplitude in time, by replacing $\epsilon $ by $\epsilon (1 - \exp(-t/T) )$ in the definition of pressure and surface height \eqref{h_ux_ux_p_def}. Practice shows that with $T = 15 \times 2 \pi$, we can keep the fast bobbing oscillations of the floater motion at low amplitude while capturing the slow dynamics induced by the wave motion.

\begin{figure}
\begin{center}
\includegraphics[scale=0.5]{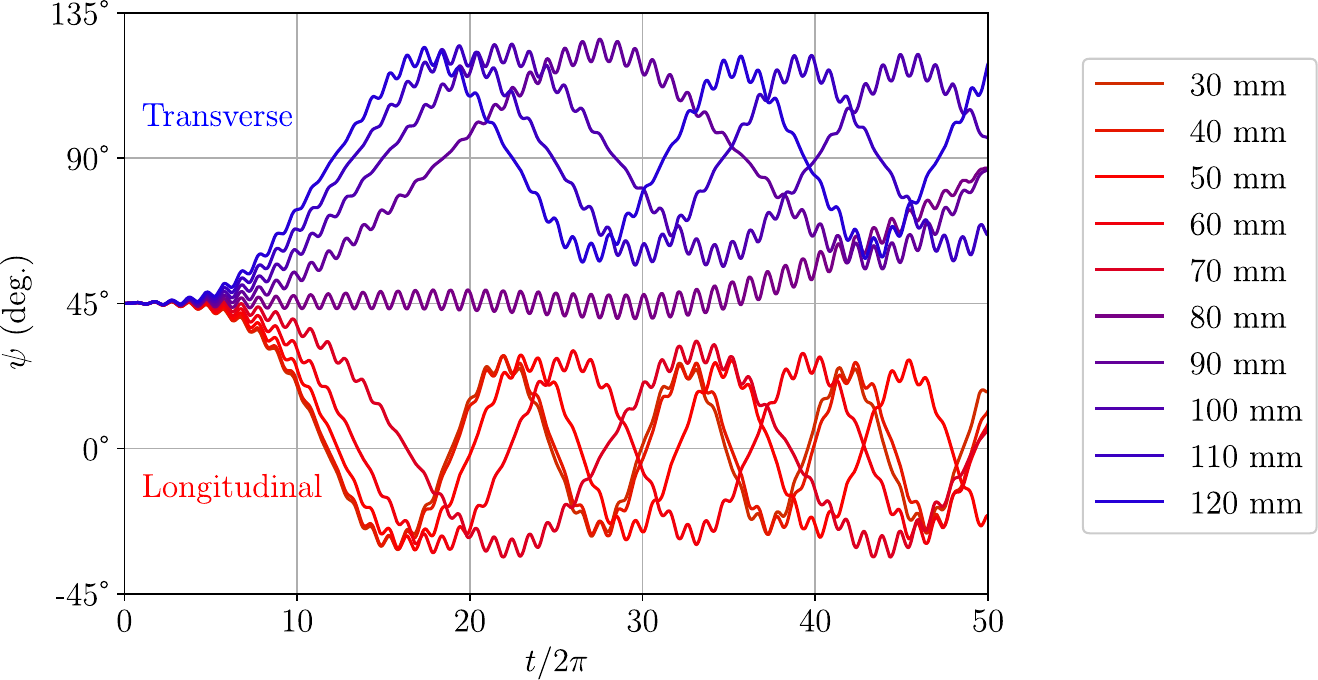}
\end{center}
\caption{Time series for yaw angle for numerically simulated floaters with  $\beta=0.44$, $L_y = 10$~mm, $L_z=5$~mm and $L_x$ varying from $30$ to $120$~mm. The incoming wave has wavelength $\lambda=0.29$~m and $\epsilon= ka = 0.16$, as in the experiments of figure~\ref{fig:ts}. The transition from a longitudinal to a transverse preferential orientation occurs near $L_{xc} \approx 75$~mm, in good agreement with the experiments.  \label{psi_time} }
\end{figure}

We now consider numerical solutions for the floater motion in presence of an incoming wave. We focus on the yaw angle motion. In a first series of simulations, we use the first order pressure field and surface elevation: $p = p^{(1)}$ and $\zeta=\zeta^{(1)}$. In figure~\ref{psi_time}, we show time series of the yaw angle $\psi (t)$, for the same experimental conditions used in figure~2: floater size $L_y = 10$~mm, $L_z=5$~mm and $L_x$ varying from $30$ to $120$~mm, density ratio $\beta=0.44$,  wavelength $\lambda=0.29$~m, wave steepness $\epsilon= ka = 0.16$, and initial yaw angle $\psi_0 = 45^o$. After a transient of typically 15 wave periods given by the ramp in the wave amplitude, the yaw angle $\psi$ shows fast oscillations (of period $2\pi$), that correspond to the back and forth motion of the floaters, superimposed to slow oscillations around either the longitudinal position $\psi=0^o$  for short floaters  or the transverse position  $\psi=90^o$ for long floaters. The fast oscillations were not visible in figure~\ref{fig:ts}(a) where the yaw angle measurement was synchronised at the wave frequency. The slow oscillations reflect the marginal stability of the fixed points in this dissipationless model. Apart from these differences, the numerical curves show a transition from longitudinal to transverse preferential orientation, somewhere in between $L_x = 70$~mm and $80$~mm, in excellent agreement with the experimental transition estimated at $L_{xc} = 75$~mm for this set of floaters.

\begin{figure}
\begin{center}
\includegraphics[width=0.7\textwidth]{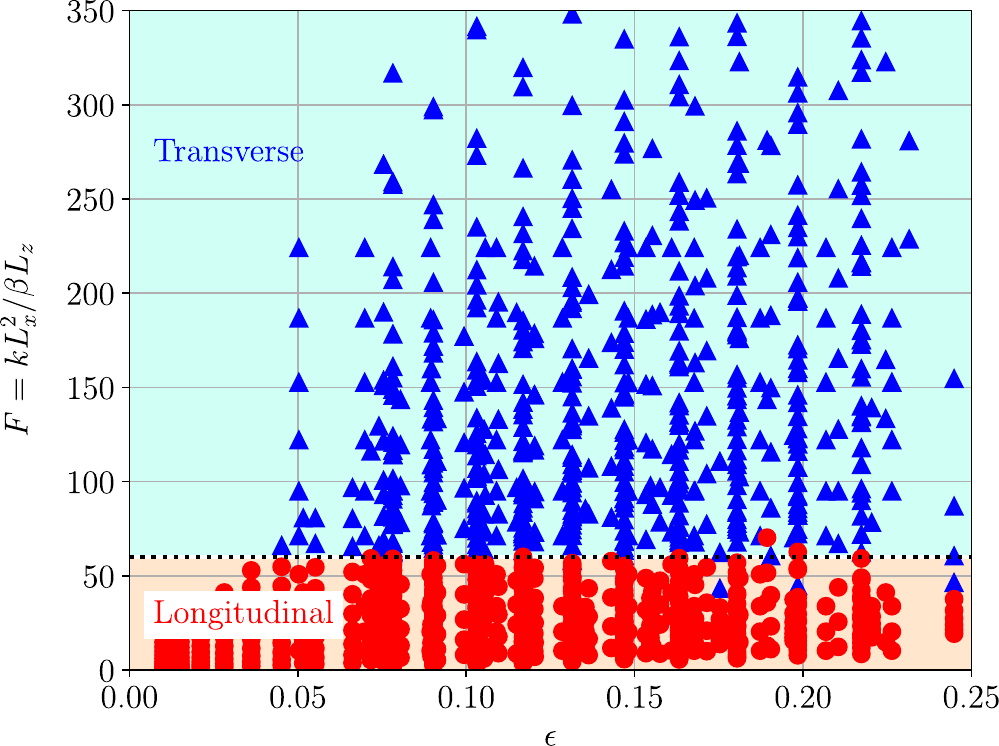}    
\end{center}
\caption{Preferential state of orientation in the $(F,\epsilon)$ plane according to the numerical simulations. More than 500 numerical simulations have been performed for varying $\beta, \delta_x,\delta_y, \delta_z, \epsilon $, showing an excellent agreement with the theoretical prediction $F_c=60$. 
\label{fig:simu_crit}
}
\end{figure}

We have done similar simulations with first order pressure  $p = p^{(1)}$ and surface elevation $\zeta=\zeta^{(1)}$ for more than 500 floaters, with parameters varied in broad ranges $\beta \in [0.03,0.97], \delta_x \in [0.004,5.43],\delta_y \in [0.01, 0.36], \delta_z \in [0.007,0.29], \epsilon \in [0.01,0.24]$.  We summarize the preferential floater orientation in figure~\ref{fig:simu_crit} using the same  representation $(F,\epsilon)$ as in the experiments (figure \ref{fig:diag}). The transition from longitudinal to transverse orientation is located at $F_c = 60$ with no dependence in $\epsilon$. Near the transition $F_c$, we observe {some} outlying data-points.  This is not entirely unexpected as a narrow interval of bistability exists near the transition, as we show below. At high $\epsilon$, higher order effects, neglected in the theory, may also become important.

\begin{figure}
\begin{center}
\includegraphics[width=\textwidth]{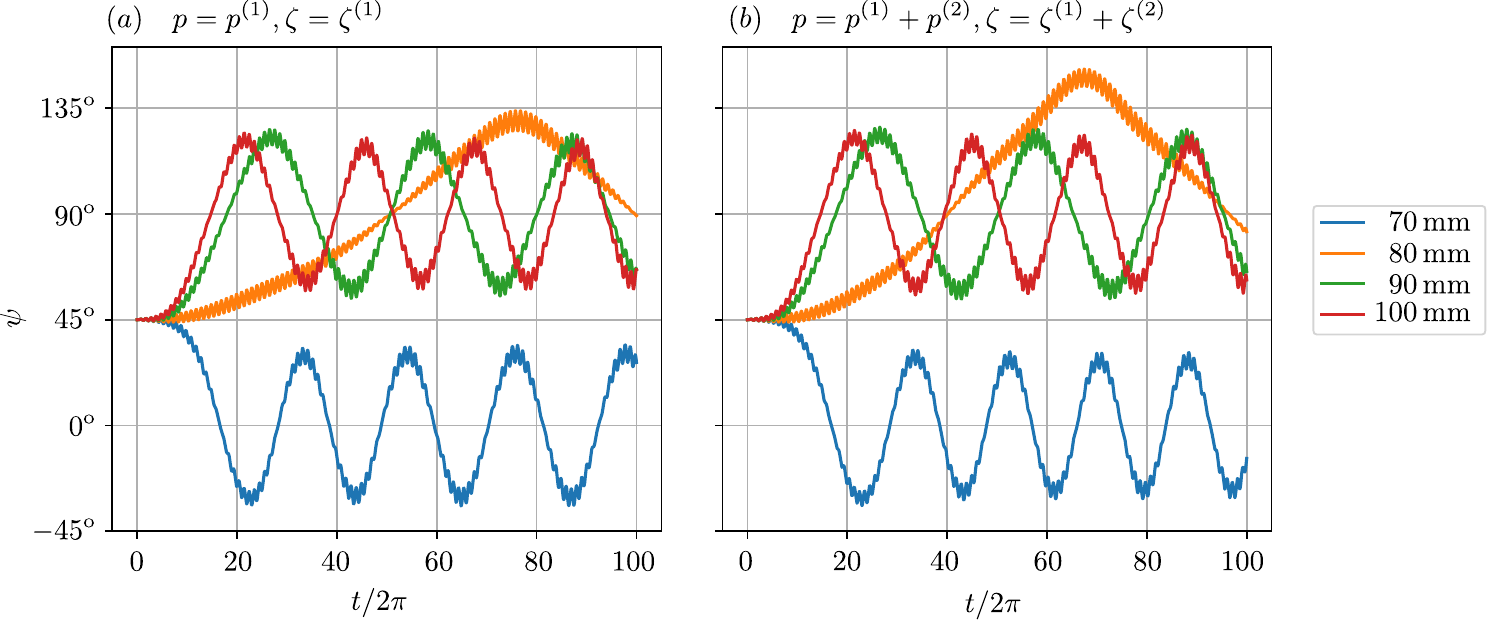} 
\end{center}
\caption{Time series for the yaw angle $\psi(t)$ of numerically simulated floaters with varying length $L_x$ (see Fig. \eqref{psi_time} for details of the floaters and the wave), with (a) first order pressure $p^{(1)}$ and surface elevation $\zeta^{(1)}$ and (b) second order corrections  $\zeta^{(2)}$ and $p^{(2)}$ also included. Preferential orientation is not much influenced by the second order part of the Stokes wave. 
\label{fig:simu_NL}
}
\end{figure}
The slow rotation towards a preferential state of orientation is a second order effect. Hence, it is legitimate to ask whether we still get the same results when the second order correction of the Stokes wave is included, when $p = p^{(1)}+p^{(2)}$ and $\zeta=\zeta^{(1)}+\zeta^{(2)}$ in the simulations. In figure~\ref{fig:simu_NL}, we compare time series of $\psi (t)$ with the first order approximation (a)  and the  second order approximation (b),  for the same parameters as in figure  \ref{psi_time}. Both figures are nearly the same, a weak difference is only observable near the transition. This indicates that second order corrections $ p^{(2)}$ and $\zeta^{(2)}$ do not have a dominant impact on preferential orientation and may in fact be ignored. This will also be theoretically justified in the next section.

\section{Asymptotic description of floater motion} 
\label{sec:theory}

In this section we use asymptotic theory to describe the motion of the floater in the limit of small wave slope $\epsilon \ll 1$ and for small floaters $\delta \ll 1$. In this theory we admit that $\delta_z \ll \delta_x$ and $\delta_z \ll \delta_y$ and we keep the aspect ratio $\delta_y/\delta_x$ arbitrary.

\subsection{Small $\epsilon$ expansion and equations of motions}

We expand the motion as a perturbative series in powers of $\epsilon$ and denote
\be
\left \{ \begin{array}{rcl}
x_c &=& \ovl{x}_{c} (\tau)  + x_c' (t)  +  O (\epsilon^2) \\
y_c &=& \ovl{y}_{c}  \\
 z_c &=& \ovl{z}_{c} + z_c' (t)  +  O (\epsilon^2)  
\end{array} \right . \quad, \quad \left \{ \begin{array}{rcl}
\varphi &=& 0  + \varphi'  (t) + O (\epsilon^2) \\
\theta& = & 0  + \theta'  (t) + O (\epsilon^2) \\
 \psi &=&\ovl{\psi}(\tau) +  \psi' (t)+ O (\epsilon^2) .
\end{array} \right .
\ee 
At leading order $O(1)$, we recognise the equilibrium positions (barred variables). We admit that horizontal positions $ \ovl{x}_{c} (\tau)$ and mean yaw angle $\overline{\psi} (\tau)$  can vary slowly, on long time-scales $\tau$. The mean position $\ovl{y}_{c}$ cannot change in our Froude-Krylov model and is of no further concern. At  order $O(\epsilon)$, we perturb the mean position with first order deviations (primes) in position  $x_c' (t),  z_c'(t)$ and angle $\psi'(t),\theta' (t), \varphi(t)$. These variables carry the harmonic response of the floater to the incoming wave. The main objective is to compute the second order mean yaw moment that leads to the evolution equation \eqref{eq:psidd} for the slow motion of the yaw angle $\ovl{\psi} (\tau)$. We do not describe the second order slow  drift of $\ovl{x}_{c} (\tau)$.

The roll and pitch angles $\varphi  ,\theta$ are small and this allows to linearise all the dependencies on these angles: $s_\theta \approx \theta' + O(\epsilon^3) , c_\theta \approx 1+ O(\epsilon^2) , s_\varphi \approx \varphi' + O(\epsilon^3) , c_\varphi \approx 1 +  O(\epsilon^2) $. The transform formula \eqref{mante_to_lab}  reduces to
\be \label{mante_to_lab_lin}
\left [ \begin{array}{c}
\bs{e}_x \\ \bs{e}_y  \\ \bs{e}_z  
 \end{array} \right ]  = \underbrace{ \left [ \begin{array}{ccc} 
 c_\psi  &  -  s_\psi  & \left ( c_\psi \theta' + s_\psi \varphi' \right ) \\
s_\psi &  c_\psi  & \left (  s_\psi \theta' - c_\psi \varphi' \right )  \\
-\theta' &  \varphi' & 1
 \end{array} \right ] }_{\mc{R}^T} 
  \left [ \begin{array}{c}
\wt{\bs{e}}_x \\ \wt{\bs{e}}_y  \\ \wt{\bs{e}}_z    
 \end{array} \right ]  + O (\epsilon^2),
\ee
and the inverse transform is defined with the transposed matrix $\mc{R}$ up to $O(\epsilon^2)$. 

We write a preliminary version of the evolution equations to identify the force and torque components that need to be calculated. The linearised evolution equations for the first order deviations $x_c', z_c' , \varphi' , \theta'$  are
\be \label{eq_first}
m \ddot{x}_{c}' = F_x'  \quad , \quad m \ddot{z}_{c}' = F_z' \quad , \quad \wt{I}_{xx} \ddot{\varphi}'  = \wt{K}_x' \quad , \quad \wt{I}_{yy} \ddot{\theta}'  = \wt{K}_y' .
\ee
Here, the forces $F_x', F_y'$ and moments $\wt{K}_x', \wt{K}_y'$ contain the $O(\epsilon)$, oscillatory part of the total force and moment. In the evolution equation for the yaw angle $\psi$ we keep $O(\epsilon)$ and $O(\epsilon^2)$ terms, since we describe both the rapid motion of $\psi'$ and the slow motion of $\ovl{\psi}$. From equations~\eqref{kinematics} and \eqref{dynamics} and using $\wt{\Omega}_x \approx \dot{\varphi}' , \wt{\Omega}_y \approx \dot{\theta}'$  we  get 
\be \label{eq_qlin_psi}
 \wt{\Omega}_z = \dot{\psi}  - \varphi'  \, \dot{\theta}' + O (\epsilon^3) \quad , \quad  \wt{I}_{zz}  \dot{\wt{\Omega}}_z  =   {\wt{K}}_{z}  +  (  \wt{I}_{xx} -  \wt{I}_{yy} ) \dot{\varphi}' \dot{\theta}' 
\ee
up to order  $O(\epsilon^2)$. Elimination of $ \wt{\Omega}_z $ gives 
\be   \label{eq_qlin_psi_2}
\wt{I}_{zz}  \dot{\psi}   =  {\wt{K}}_{z}  + \frac{d}{dt} \left(  \wt{I}_{zz}     \varphi'  \, \dot{\theta}' \right )   + (  \wt{I}_{xx} -  \wt{I}_{yy} ) \dot{\varphi}' \dot{\theta}' .
\ee
We now remark the following simplification. From the transform \eqref{mante_to_lab_lin} and using $\wt{K}_x = \wt{K}_x' + O(\epsilon^2)$, $\wt{K}_y = \wt{K}_y' + O(\epsilon^2)$ and the evolution equations \eqref{eq_first}, we get    
\ba
K_z
& \approx & - \theta' \wt{K}_x' + \varphi' \wt{K}_y' + {\wt{K}}_{z} + O(\epsilon^2)  \nonumber \\
&  \approx &  - \theta' \wt{I}_{xx} \ddot{\varphi}'   + \varphi'  \wt{I}_{yy} \ddot{\theta}'   + {\wt{K}}_{z} + O(\epsilon^2) .
\ea
We reorganise this equation to isolate the component ${\wt{K}}_{z} $ and substitute this into  \eqref{eq_qlin_psi_2}: 
\ba
 \wt{I}_{zz}  \ddot{\psi} &=& 
K_z +  \underbrace{\frac{d}{dt} \left (   \wt{I}_{xx}    \dot{\varphi}'  \theta'   -    \wt{I}_{yy}   \varphi'   \dot{\theta}'  +     \wt{I}_{zz}   \varphi'  \, \dot{\theta}'   \right ) }_{ \text{second order harmonics} \ \sim \  e^{\pm i 2 t}} + O (\epsilon^3 ).
\ea
In the right hand side, we find $K_z$, the vertical moment component in the {\it laboratory frame}, next to a term $d/dt (\ldots) $ that is a time-derivative of  $O(\epsilon^2)$ products. This second order term has vanishing time-average, so it cannot affect $\ovl{\psi}$. Hence, we can derive the equations for both yaw angle variables $\ovl{\psi}$ and $\psi'$ from the simpler balance $ \wt{I}_{zz}  \ddot{\psi} \approx K_z$. Separating the yaw moment in a rapidly varying $O(\epsilon)$ part $K_z'$ and  a mean yaw moment $\ovl{K}_z$ of order $O(\epsilon^2)$, we obtain 
\ba \label{eq_psi}
\wt{I}_{zz}  \ddot{\psi}'      =   K_z'  \quad , \quad  \wt{I}_{zz}  \ddot{\ovl{\psi}}      =  \overline{K}_z 
\ea
as preliminary evolution equations for the yaw motion. 

\subsection{Theoretical calculation of force and moment components}

To calculate the required force and moment components, we use an alternative formulation of equation \eqref{FpKp} in terms of the flow field $\bs{u}$ and with volume integrals. We replace $d \bs{S}= - d \bs{S}_{ext}$ in equation~\eqref{FpKp}, so that the surface element of the submerged volume points points towards towards the fluid. Using then the divergence theorem on the interior of $S_{sub}$ and Euler's equation, $\pd_t \bs{u}  +( \bs{u} \cdot \bnabla )\bs{u} + \bs{e}_z = - \bnabla p $, we get that
\bse  \label{FpKp_alt}
\ba
\bs{F} &= &  
\int_{V_{sub}} ( \pd_t  \bs{u}  +  ( \bs{u} \cdot \bnabla )\bs{u}  )  \, dV  + \left (   \int_{V_{sub}}   \, dV  - m \right ) \bs{e}_z \\
\bs{K} &=&   \int_{V_{sub}} ( \bs{r} - \bs{r}_c) \times    ( \pd_t  \bs{u} +  ( \bs{u} \cdot \bnabla )\bs{u}  )   \, d V + \int_{V_{sub}} ( \bs{r} - \bs{r}_c) \times  \bs{e}_z   \,  d V
\ea
\ese
 in the Froude-Krylov approximation, with $V_{sub}$ the submerged volume. We now inject the flow-velocity $ \bs{u} =  \bs{u}^{(1)} + O (\epsilon^3)$ as defined by equation  \eqref{h_ux_ux_p_def} and introduce two approximations of the submerged volume: \\
 \begin{itemize}
 \item[$$] $ V_{sub}^{(0)} $ :  equilibrium submerged volume  
 \item[$$]  $ V_{sub}^{(0+1)} $ :  following the floater through its first order motion, with surface at $z=\zeta^{(1)} $. \\
 \end{itemize}
We can integrate over both these volumes and also introduce the notation
 \be
  \int_{V_{sub}^{(1)}}  \Big (  \ldots   \Big )   dV   = \int_{V_{sub}^{(0+1)}}  \Big (  \ldots   \Big )   dV -   \int_{V_{sub}^{(0)}}  \Big (  \ldots   \Big )   dV .
 \ee
 This isolates the part in the integrals that is entirely due to the wave or floater motion. At leading order $O(1)$, without flow, we have the Archimedes balance
  \be
 \int_{V_{sub}^{(0)}}   \, dV  - m = 0 , \quad \int_{V_{sub}^{(0)}} ( \bs{r} - \bs{r}_c) \times  \bs{e}_z   \,  d V =\bs{0} .
 \ee
These equations can be used to find the equilibrium state \eqref{eq:equilibrium}. For the first order forces and moments, we have  
\bse  \label{FpKp_alt_prime}
\ba
\bs{F}' &= & \int_{V_{sub}^{(0)}}  \pd_t  \bs{u}^{(1)}  \, dV  +  \int_{V_{sub}^{(1)}}    \, dV   \bs{e}_z \\
\bs{K}' &=&   \int_{V_{sub}^{(0)}} ( \bs{r} - \bs{r}_c) \times     \pd_t  \bs{u}^{(1)} + \int_{V_{sub}^{(1)}} ( \bs{r} - \bs{r}_c) \times  \bs{e}_z   \,  d V.
\ea
\ese
For the mean yaw moment, we need a second order approximation. Using an overbar to indicate the time-average over the short time-scale $t$, we have 
 \be \label{Kz_mean}
\ovl{K}_z  =  - \ovl{\int_{V_{sub}^{(0+1)}} (y-y_c) \, \pd_t u_x^{(1)}  \, d V} - \ovl{ \underbrace{\int_{V_{sub}^{(0)}} (y-y_c) \, (\bs{u}^{(1)}\cdot \bnabla  ) u_x^{(1)}   \ d V }_{ =0 } } + O (\epsilon^3).
\ee  
The advective term vanishes on average because $(\bs{u}^{(1)}\cdot \bnabla  ) u_x^{(1)} \sim  \sin 2(x-t)$. We remark that no second order wave characteristics appear in this mean yaw moment formula. This explains why second order wave characteristics are not so important in the problem of preferential orientation, as we have seen in figure \ref{fig:simu_NL}.

The integrals  in equations \eqref{FpKp_alt_prime} and \eqref{Kz_mean} cannot be exactly calculated, but we can obtain asymptotic approximations in the small floater limit $\delta \ll 1$. The general idea is to use Taylor series to replace  the flow field and surface elevation in the vicinity of the small floater with polynomial approximations that are more easily integrated. We only present the main methods,  details are in the supplementary material. 

We parametrise the submerged volume  in floater frame coordinates: 
\be
V_{sub} \ : \ \wt{x} \in [-\delta_x/2, \delta_x/2] , \  \wt{y} \in [-\delta_y/2, \delta_y/2] , \  \wt{z} \in [-\delta_z/2, \wt{\zeta} (\wt{x},\wt{y},t)] .
\ee
Here $\wt{\zeta} (\wt{x},\wt{y},t)$ represents the height of the water surface, in the $\wt{z}$-direction, as seen from $C$. We implicitly suppose that the top face of the floater is never partially submerged ($\wt{\zeta} < \delta_z/2)$ and that the bottom face is always totally submerged ($\wt{\zeta} > -\delta_z/2)$. This is a sensible approximation for low $\epsilon$ and intermediate values of $\beta$. To find a polynomial approximation of $\wt{\zeta}$, we write a Taylor expansion of $\zeta (x,t)$ around $x_c$:
\be \label{h_taylor} 
\zeta(x,t) = \zeta_c  + (x-x_c )  \pd_x \zeta_c +   \frac{(x-x_c )^2}{2} \pd^2_{xx} \zeta_c +  \frac{(x-x_c )^3}{6} \pd^3_{xxx} \zeta_c   + \ldots
\ee 
The index ${}_c$ is used to express that the field is evaluated at the center of the floater $C$, e.g.  $\zeta_c = \zeta (x_c,t) = \epsilon \sin (x_c -t) $, $\pd_x \zeta_c = \pd_x \zeta |_{x=x_c} = \epsilon \cos (x_c -t)$, etc. We inject this expansion in the equation  $z = \zeta(x,t) $ that defines the free surface and we replace both $z \approx z_{c}  + ( - \theta' \wt{x} + \varphi'  \wt{y} + \wt{z} ) $ and $x-x_c \approx c_\psi \wt{x} - s_\psi \wt{y} + O(\epsilon)$, the leading order parts of the coordinate transform formula \eqref{mante_to_lab_lin}.
Reorganising this into $\wt{z} = \wt{\zeta}(\wt{x},\wt{y}, t) $ gives 
\ba
\wt{\zeta}(\wt{x},\wt{y}, t)   &=& -\ovl{z}_{c}    - z_c'  +  \zeta_c  + ( \theta'  + c_\psi \pd_x \zeta_c ) \wt{x} +  ( -\varphi'  - s_\psi \pd_x \zeta_c ) \wt{y}    \label{htdef}  \\
&  & + \frac{1}{2} \left ( c_\psi \wt{x} - s_\psi \wt{y} \right )^2   \pd^2_{xx} \zeta_c  + \frac{1}{6} \left ( c_\psi \wt{x} - s_\psi \wt{y} \right )^3   \pd^3_{xxx} \zeta_c   + \, O (\epsilon^2, \epsilon \delta^4) . \nonumber
\ea
The free surface is indeed $ -\ovl{z}_{c} $ above $C$ at equilibrium. The wave causes $O(\epsilon)$ corrections that here are separated in a local uniform elevation, linear inclination and quadratic and cubic corrections. The integral over the equilibrium submerged volume is defined as 
\ba
\int_{V_{sub}^{(0)}}  \Big (  \ldots   \Big )   dV  &=&   \int_{-\delta_x /2}^{\delta_x/2}   \int_{-\delta_y/2}^{\delta_y/2}   \int_{-\delta_z/2}^{ - \ovl{z}_c  }  \Big (  \ldots   \Big )   \Big  |_{( \ovl{x}_c,\ovl{z}_c,  \ovl{\psi})}   d \wt{x} \,d \wt{y} \, d \wt{z} .  
\ea
The notation $|_{( \ovl{x}_c,\ovl{z}_c,  \ovl{\psi})}$ suggests that we need to replace $x_c=\ovl{x}_c, z_c=\ovl{z}_c, \psi = \ovl{\psi}$ in the integrand because we need to evaluate it at the equilibrium position. The integral over the perturbed submerged volume  $V_{sub}^{(0+1)}$  is more complex: we need to integrate up until the surface deformed by the wave and we also need to evaluate the integrand at the first order perturbed floater position $x_c=\ovl{x}_c+x_c', z_c=\ovl{z}_c+z_c', \psi = \ovl{\psi}+ \psi'$. We have
\ba
&& \int_{V_{sub}^{(0+1)}}  \Big (  \ldots   \Big )   dV  \nonumber \\
&& \quad \approx     \int_{-\delta_x /2}^{\delta_x/2}   \int_{-\delta_y/2}^{\delta_y/2}   \int_{-\delta_z/2}^{ \wt{\zeta}(\wt{x},\wt{y}, t) } \Big (  \ldots   \Big )   \Big    |_{(\ovl{x}_c+x_c',\ovl{z}_c+z_c', \ovl{\psi}+\psi')}    d \wt{x} \,d \wt{y} \, d \wt{z}     \   \nonumber \\
 \nonumber \\
&& \quad  \approx  \underbrace{ \int_{-\delta_x /2}^{\delta_x/2}   \int_{-\delta_y/2}^{\delta_y/2}   \int_{-\delta_z/2}^{ - \ovl{z}_c  } \left (1 +  x_c' \frac{\pd}{\pd x_c}  + z_c'  \frac{\pd}{\pd z_c}  + \psi' \frac{\pd}{\pd \psi} \right  )    \Big (  \ldots   \Big )   \Big |_{(\ovl{x}_c,\ovl{z}_c, \ovl{\psi})}    d \wt{x} \,d \wt{y} \, d \wt{z}           }_{\text{equilibrium and deviation due to motion $x_c',z_c',\psi'$ }} \nonumber \\
&& \quad\quad \quad + \underbrace{ \int_{-\delta_x /2}^{\delta_x/2}   \int_{-\delta_y/2}^{\delta_y/2}  \left ( \wt{\zeta}(\wt{x},\wt{y},t) + \ovl{z}_c   \right )  \Big ( \ldots  \Big )   \Big  |_{(\ovl{x}_c,\ovl{z}_c, \ovl{\psi})}     d \wt{x} \,d \wt{y}  }_{\text{deviation due to locally varying submersion}}  \ + \ O (\epsilon^2) . \label{Vsub01def}
\ea
 This formula is correct up to order  $O(\epsilon)$ and requires some explanations. First of all, we have split the $\wt{z}$-integral in two parts, one over the equilibrium submersion interval $\wt{z} \in [-\delta_z/2,-\ovl{z}_c]$ and one over the $O(\epsilon)$ deviation $\wt{z} \in [-\ovl{z}_c, \wt{\zeta}]$.  In the first part, we use a Taylor expansion $(1+ x_c' \pd_{x_c} + ... )$ to re-express everything at the equilibrium position $x_c=\ovl{x}_c, z_c=\ovl{z}_c, \psi = \ovl{\psi}$. In the second part, we can simplify the integration over $\wt{z}$ as the integration interval has a size $ \wt{\zeta}+ \ovl{z}_c$ that is of order $O(\epsilon)$. 

Our choice to parametrise the integrals in floater frame coordinates requires that we also express the integrands in floater frame coordinates. We have $\bs{r}-\bs{r}_c = \wt{x} \wt{\bs{e}}_x + \wt{y} \wt{\bs{e}}_y + \wt{z} \wt{\bs{e}}_z $ and the coordinate transform \eqref{mante_to_lab_lin}. We use Taylor expansions around the floater center $\bs{r}_c$ and replace all fields with polynomial approximations. Denoting in short $\bs{a}= \pd_t \bs{u}^{(1)}$ the local fluid accceleration, we have 
\ba
\bs{a} &=& \bs{a}_{c}  +  (x-x_c) \pd_x \bs{a}_{c} + (z-z_c) \pd_z \bs{a}_{c}  \\
& & + \frac{1}{2}  (x-x_c)^2 \pd_{xx}^2 \bs{a}_{c} + \frac{1}{2}   (z-z_c)^2 \pd^2_{zz} \bs{a}_{c}  +    (x-x_c) (z-z_c)\pd^2_{xz} \bs{a}_{c}   
+ O (\epsilon \delta^3) . \nonumber
\ea
We substitute the transform $x-x_c = c_\psi \wt{x} - s_\psi \wt{y}     +  \left ( c_\psi \theta' + s_\psi \varphi' \right ) \wt{z}  $ and $z-z_c = - \theta' \wt{x} + \varphi'  \wt{y}  +  \wt{z}  $, keeping the order  $O(\epsilon)$ terms that are proportional to $\theta'$ and $\varphi'$. This produces a lengthy expression detailed in the supplementary material that can be summarized as 
\ba
\bs{a} &=&   \bs{a}^{(1)}  +   \bs{a}^{(2)}  + O (\epsilon^3) .
\ea
Next to the leading  $O(\epsilon)$ term $ \bs{a}^{(1)}$, there is a $O(\epsilon^2)$ deviation  $\bs{a}^{(2)}$ that is not related to $\bs{u}^{(2)} = \bs{0}$, but rather due to the fact that angular oscillations $\theta', \varphi'$ cause the floater to feel spatial variations of the flow-speed.

Having explained the main mathematical methods, we can now write explicit formula for the first order force and torque components  
\bse
\ba 
F_x' &=&    \int_{V_{sub}^{(0)}} a_x^{(1)}  dV \,    \label{Fx_new} \\ 
F_z' &= &    \int_{V_{sub}^{(0)}} a_z^{(1)}  \,  d V +   \int_{V_{sub}^{(1)}}   d V      \label{Fz_new} \\
\wt{K}'_x &=&    \int_{V_{sub}^{(0)}}   \left (   \wt{y}     a_z^{(1)}   +  s_\psi  \wt{z}  a_x^{(1)}   - \wt{z}   \varphi'   \right ) dV +  \int_{V_{sub}^{(1)}}   \wt{y}    d V    \label{wtKx_new}\\
\wt{K}'_y &=&   \int_{V_{sub}^{(0)}}  \left ( -  \wt{x}     a_z^{(1)}   +c_\psi  \wt{z}   a_x^{(1)}  - \wt{z}   \theta'   \right ) dV +  \int_{V_{sub}^{(1)}}  (- \wt{x}   )  d V   \label{wtKy_new} \\
K'_z  &=& 
-  \int_{V_{sub}^{(0)}}  \left (  s_\psi  \wt{x}    +  c_\psi  \wt{y}   \right ) a_x^{(1)}  dV .
\ea
\ese
For the second order, mean yaw moment, we have 
\ba
 \ovl{K}_z &=&   - \ovl{ \int_{V_{sub}^{(0)}}   \left (  s_\psi \theta' - c_\psi \varphi' \right )  \wt{z}  a_x^{(1)} d V }    \    -  \ovl{  \int_{V_{sub}^{(0)}}  \left (  s_\psi  \wt{x}    +  c_\psi  \wt{y}   \right ) a_x^{(2)}  d V }     \nonumber \\ 
 & &   -  \ovl{ \int_{V_{sub}^{(1)}}  \left (  s_\psi  \wt{x}    +  c_\psi  \wt{y}   \right ) a_x^{(1)}  dV. }  \label{Kz2_new}
\ea
The calculation of these integrals is long but straightforward and in the supplementary material, we provide some details. One practical question that appears is: to which order in non-dimensional floater size $\delta$ do we need to push the Taylor series? Considering the equations of motion  \eqref{eq_first} and \eqref{eq_psi}, and the fact that  $m= O(\delta^3)$ and  $\bs{I}= O(\delta^5)$, we have calculated
\ba
 F'_x, F'_z & & \ \  \text{up to order} \  O(\epsilon \delta^4)  \nonumber  \\
\wt{K}'_x, \wt{K}'_y, K_z' & &  \ \  \text{up to order} \  O(\epsilon \delta^6)  \label{order_delta}\\
\ovl{K}_z &&\ \   \text{up to order} \ O(\epsilon^2 \delta^6). \nonumber 
\ea
This is necessary to access to the leading, $O(\epsilon)$ description of $x_c', z_c', \varphi', \theta', \psi'$, $O(\epsilon^2)$ description of $\ovl{\psi}$, but also to floater shape-related corrections of respective orders $O(\epsilon \delta)$ and $O(\epsilon^2 \delta)$. In the limit $\delta \ll 1$, these shape corrections seem much smaller but this is without considering that these $O(\epsilon \delta)$ and $O(\epsilon^2 \delta)$  terms are actually taking the form of $O(\epsilon\delta_x^2/\beta \delta_z ) $ and $O(\epsilon^2\delta_x^2/\beta \delta_z ) $ terms. For strongly elongated floaters, we can have 
\be
F = \frac{\delta_x^2}{\beta \delta_z} \gg 1 
\ee
and in that case, the shape related corrections become larger than the so-called leading order terms. This is precisely what happens when the floaters change their preferential orientation. In practice, we really need to push calculation to the orders  \eqref{order_delta}, but we find that this results in very long formula. This motivated us to add one extra assumption. In the experiment, all the floaters are always thinner in the $z$-direction and this means that 
\be \label{flat_floater}
\delta_x , \delta_y \gg \delta_z
\ee
is a fair assumption. By exploiting this information in the calculation of the forces and moments we can ignore  (i) all $O(\delta_z)$ terms with respect to $O(1)$ terms,  (ii) all $O(\delta_z^2)$ terms with respect to $O(\delta_x^2, \delta_y^2)$ terms.  This greatly simplifies the resulting formula for the forces and moments and produces a physically relevant result. We can in fact even further simplify the model by taking into account that our floaters are also very elongated $\delta_x \gg \delta_y$ in our experiments, but this will be done at the end. 

\subsection{Results}
\label{sec:res}

We now present the result of the asymptotic calculation. With the first order force and moments $F_x', F_z', \wt{K}_x', \wt{K}_y', {K}_z'$ calculated, we write the differential equations for the first order motion $x_c',z_c',\varphi', \theta', \psi'$ and  solve them. This yields the harmonic response 
\bse \label{solution_first}
\ba
{x}'_c  & \approx &       \epsilon   \cos (\ovl{x}_c -t )  \\
z_c' &\approx & \epsilon \left ( 1 - \frac{1}{24} \left (\ovl{c}^2_\psi \delta_x^2 + \ovl{s}_\psi^2 \delta_y^2  \right  )  \right ) \sin (\ovl{x}_c-t) . \label{zcprime} \\
\varphi' &\approx &- \epsilon \ovl{s}_{{\psi}}   \left (1 -  \left (  \ovl{s}_{{\psi}}^2  \dfrac{\delta_y^2}{40}   + \ovl{c}_{{\psi}}^2  \dfrac{\delta_x^2 }{24} \right )     \right ) \cos (\ovl{x}_c -t )   \\
\theta' &\approx& - \epsilon \ovl{c}_{{\psi}}  \left ( 1-    \left (  \ovl{c}_{{\psi}}^2  \dfrac{\delta_x^2}{40}   + \ovl{s}_{{\psi}}^2  \dfrac{\delta_y^2 }{24} \right )   \right ) \cos (\ovl{x}_c -t ) \label{thetaprime}  \\
\psi' &\approx &   \epsilon  \left ( \frac{\delta_x^2- \delta_y^2}{\delta_x^2+ \delta_y^2} \right )   \ovl{s}_{{\psi}}  \ovl{c}_{{\psi}}   \sin (\ovl{x}_c - t).  \label{psi_prime}
\ea
\ese
At leading $O(\epsilon)$ order, we find in $x_c'$ and $z_c'$ that the floater oscillates around its mean position, just as a material particle on the wave-surface would do. Although this is less trivial to see, the leading $O(\epsilon)$ expressions for the angles $\varphi', \theta'$ are such that the floater rotates so to align with the local wave slope (imagine the rotation of a surfboard on a long wavelength wave that passes). The yaw angle $\psi'$ oscillates more for elongated floaters with $\delta_x \gg \delta_y$ and for $\ovl{\psi}$ close to $45^o$. Next the leading $O(\epsilon)$ terms, we also have some smaller, shape-related corrections of order $O( \epsilon \delta_x^2, \epsilon \delta_y^2) $ in the $z_c', \varphi', \theta'$ variables. These corrections
may seem utterly small but in particular the $O( \epsilon \delta_x^2)$ correction in $z_c'$ is essential in the theory. Physically, these corrections are due the fact that the water surface is not perfectly flat at the scale of the floater. It is this waterline curvature that induces a modification in the buoyancy force that ultimately causes long floaters to prefer the transverse equilibrium. 


The calculation of the mean yaw moment $\ovl{K}_z$ is challenging but leads to the following nonlinear equation of motion for $\ovl{\psi}$:  
\ba
\ddot{\ovl{\psi}} = \epsilon^2 \ovl{c}_\psi \ovl{s}_\psi \left [ \ovl{c}_\psi^2 \left ( - \frac{1-b}{(1+b)^2} +  \frac{F}{60} \,\, \frac{ 1  - \frac54 b}{1+b} \right ) 
- \ovl{s}_\psi^2 \left ( - \frac{1-b^{-1}}{(1+b^{-1})^2} +  \frac{bF}{60} \,\, \frac{ 1  - \frac54 b^{-1}}{1+b^{-1} } \right )  \right ], \nonumber \\
\label{psi_gen_eq}
\ea
with $b=(\delta_y/\delta_x)^2$ and $F = {\delta_x^2}/{ \beta \delta_z}$. We can rewrite this equation in conservative form, $\ddot{\ovl{\psi}}  = - \partial V / \partial \ovl{\psi}  $, by introducing the effective potential   
\ba
V(\ovl \psi) = \frac14 \epsilon^2 \left[ 
\ovl{c}_\psi^4 \left ( - \frac{1-b}{(1+b)^2} +  \frac{F}{60} \,\, \frac{ 1  - \frac54 b}{1+b} \right ) 
+ \ovl{s}_\psi^4 \left ( - \frac{1-b^{-1}}{(1+b^{-1})^2} +  \frac{bF}{60} \,\, \frac{ 1  - \frac54 b^{-1}}{1+b^{-1} } \right ) \right] . \nonumber \\
\label{veff_psi} 
\ea
We verify that this potential is invariant under an exchange of $\delta_x$ and $\delta_y$ (i.e., $b \rightarrow b^{-1}$) and a rotation of the angle $\ovl{\psi} \rightarrow \ovl{\psi}  + \pi/2$.

\begin{figure}
\begin{center}
\includegraphics[width=.9\textwidth]{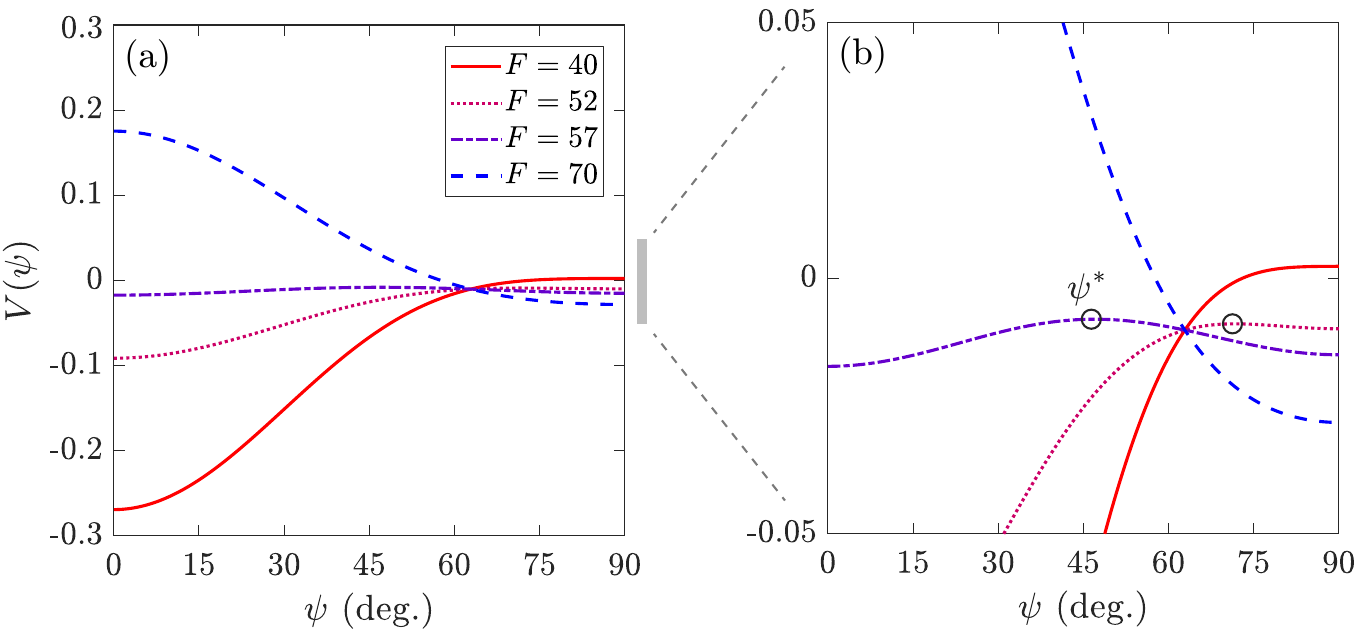}    
\end{center}
\caption{(a) Effective potential \eqref{psi_gen_eq} for a floater of aspect ratio $\delta_y/\delta_x = 0.2$. For $F=40$, the longitudinal orientation $\ovl \psi = 0^o$ is unconditionally stable, and for $F=70$ the transverse orientation $\ovl \psi = 90^o$ is unconditionally stable. The magnification in (b) highlights the unstable points $\ovl{\psi^*}$ (marked with a symbol $\circ$) separating the two bistable solutions for intermediate values of $F$.
\label{fig:psi4}
}
\end{figure}

We first examine the stable equilibria for arbitrary aspect ratio $0<b<1$. Since $V(\ovl \psi)$ is a linear combination of $\cos^4 \ovl \psi$ and $\sin^4 \ovl \psi$, it clearly admits $\ovl \psi = 0^o$ and $\ovl \psi = 90^o$ as extrema, and we can see that for $F \rightarrow 0$ the first one is a minimum and the second one a maximum, and vice-versa for $F \rightarrow \infty$. This confirms that short floaters (small $F$) tend to align longitudinally and long floaters (large $F$) transversely, in agreement with the experiments and the numerical simulations.

In the experiments and in the numerical simulations, the aspect ratio $\delta_y / \delta_x$ lies in the range $0.08-0.3$. We consider the representative intermediate value $\delta_y / \delta_x = 0.2$ to illustrate in figure~\ref{fig:psi4}(a) the potential for various values of $F$. The change of stability between $\ovl \psi = 0^o$ and $\ovl \psi = 90^o$ is clearly visible as $F$ is increased, but their basins of attraction are not identical:  the curvature of the potential well is much more pronounced around $\ovl{\psi}=0^o$ than around $\ovl{\psi}=90^o$, leading to ``faster'' slow oscillations around the longitudinal than around the transverse equilibrium. In the experimental angle trackings shown in figure~\ref{fig:ts}(a), because of the dissipation (not accounted for in our inviscid diffractionless model), these slow oscillations are rapidly damped and the floaters converge towards their stable equilibrium. However, the rapid convergence towards $\ovl{\psi}=0^o$ for short floaters, and the much slower dynamics with large erratic excursions around $\ovl{\psi}=90^o$ for longer floaters, may be consequences of these different dynamics.

\begin{figure}
\begin{center}
\includegraphics[width=.6\textwidth]{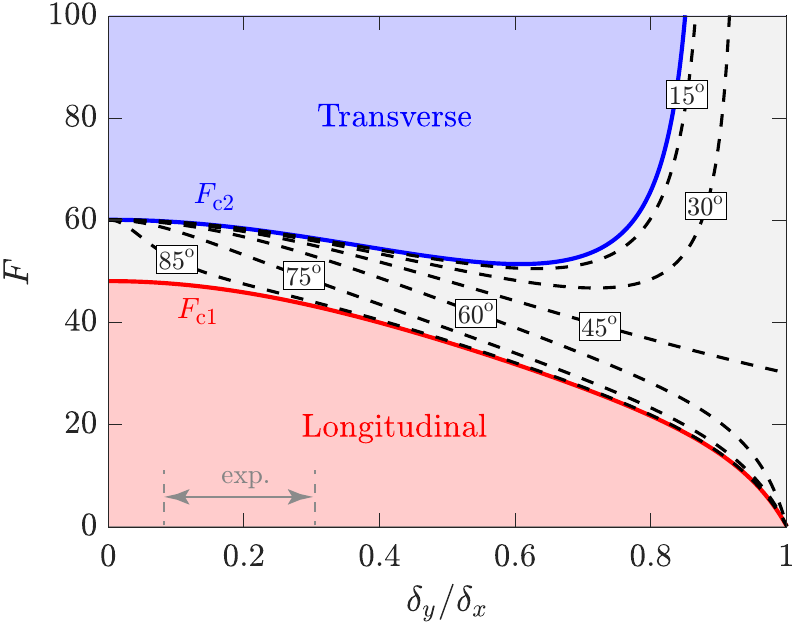}
\caption{Diagram of the equilibrium position for a floater of aspect ratio $\delta_y / \delta_x = b^{1/2}$ and parameter $F$. The red and blue areas denote the unconditionally stable equilibria $\ovl \psi = 0^o$ (longitudinal) and  $\ovl \psi = 90^o$ (transverse). The dashed lines in the intermediate bistable area show the unstable solution $\ovl{\psi^*}$ (values indicated in the boxes) given by equation~\eqref{eq:psistar}, separating the longitudinal and transverse basins of attraction. The horizontal arrow indicates the range of aspect ratio considered in the experiments.}
\label{fig:bis}
\end{center}
\end{figure}

{For an intermediate range of $F$, the potential admits a local maximum at $\ovl{\psi^*}$,
\begin{equation}
\tan^2 \ovl{\psi^*} = \frac{- \dfrac{1-b}{(1+b)^2} +  \dfrac{F}{60} \,\, \dfrac{ 1  - \frac54 b}{1+b} }{ - \dfrac{1-b^{-1}}{(1+b^{-1})^2} +  \dfrac{bF}{60} \,\, \dfrac{ 1  - \frac54 b^{-1}}{1+b^{-1} }} ,
\label{eq:psistar}
\end{equation}
as illustrated in the magnification in figure~\ref{fig:psi4}(b). This local maximum separates the two basins of attraction, indicating a bistability in the system:} for an initial condition $\ovl \psi_0 < \ovl{\psi^*}$, the yaw angle is attracted to the longitudinal equilibrium $\ovl \psi=0^o$, whereas for $\ovl \psi_0 > \ovl{\psi^*}$ it is attracted to the transverse equilibrium $\ovl \psi=90^o$. The boundaries of this bistable range are obtained by taking
$\ovl{\psi^*} = 0^o$ and $\ovl{\psi^*} = 90^o$ in equation~\eqref{eq:psistar}, yielding
\begin{equation}
F_{c1} = 60 \frac{1-b}{(1+b)(\frac54 -  b)}, \qquad F_{c2} = 60 \frac{1-b}{(1+b)(1-\frac54 b)} \quad \mbox{(for $b<4/5$)}.
\label{eq:fc12}
\end{equation}
In other words,  the longitudinal equilibrium is unconditionally stable for $F < F_{c1}$, the transverse equilibrium is unconditionally stable for $F > F_{c2}$, and both equilibria coexist in the intermediate case $F_{c1} < F < F_{c2}$, depending on how the initial angle ${\ovl \psi}_0$ compares to the separatrix $\ovl{\psi^*}$.  This is summarized in the regime diagram in figure~\ref{fig:bis}, showing the regions of unconditionally stable equilibria in red and blue, and the intermediate bistable region in gray. The set of dotted lines show the separatrix between the longitudinal and the transverse equilibrium for a given initial condition ${\ovl \psi}_0$. We note that $F_{c2} \rightarrow \infty$  and $F_{c1} \rightarrow 0$ in the limit $b \rightarrow 1$: the bistable domain extends over all values of $F$ for nearly square floaters. This means that preferential orientation of a nearly square floater is entirely governed by its initial angle $\ovl \psi_0$ and not by $F$.

\begin{figure}
\begin{center}
\includegraphics[width=.6\textwidth]{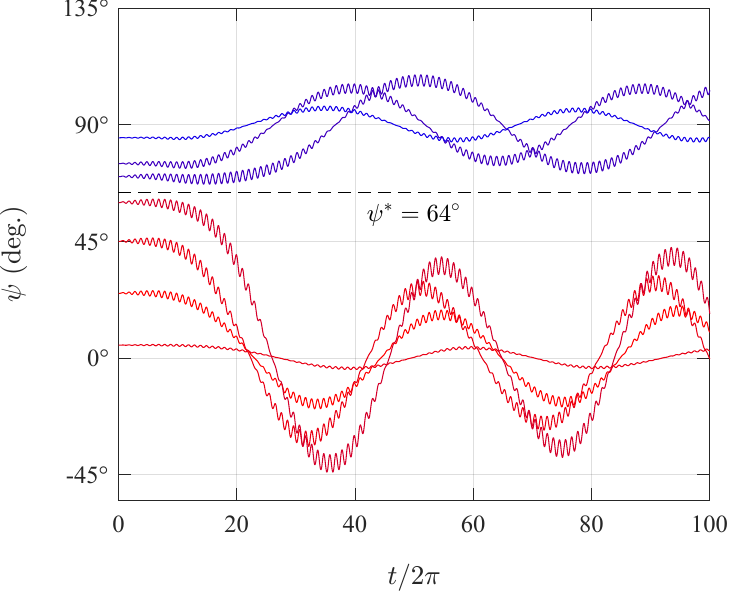}    
\end{center}
\caption{Time evolution of the yaw angle $\psi(t)$ for $F = 56$, $\delta_y/\delta_x = 0.2$, and $\epsilon=0.15$, illustrating the bistability of the longitudinal and transverse equilibria close to the transition. The separatrix here is $\ovl{\psi^*} = 64^o$, and floaters released with an initial yaw angle $\psi_0< \ovl{\psi^*}$ oscillate around the longitudinal equilibrium, while those with $\psi_0> \ovl{\psi^*}$ oscillate around the transverse equilibrium.
\label{fig:num_bistab}
}
\end{figure}

Considering again the typical experimental value $\delta_y / \delta_x = 0.2$,  the model predicts bistability for $F \in [45.8, 58.3]$, with a separation between the two stable equilibria for an initial angle $\ovl \psi_0 = 45^o$ at $F \simeq 57.7$. Although this is in overall qualitative agreement with the experiments, in particular regarding the data labeled as `indistinct' close to the transition in the regime diagram in figure~\ref{fig:diag}, the experimental uncertainties ($\pm 15^o$ for $\ovl \psi_0$ and $\pm 20\%$ for $F$) make a systematic exploration of the bistable regime difficult. On the other hand, the bistability can be tested numerically. Figure~\ref{fig:num_bistab} shows the time evolutions of $\psi(t)$ for floaters with $F=56$ released at various initial angles $ \psi_0$ ranging from 0 to $90^o$. The curves clearly separate in two groups, with $\psi(t)$ oscillating around $0^o$ for $\psi_0 \leq 60^o$ and around $90^o$ otherwise, in excellent agreement with the predicted separatrix at $\ovl{\psi^*} = 64^o$.

The limit of very thin floaters, $\delta_y / \delta_x \rightarrow 0$, provides an interesting simplification of the problem. In this limit, the boundaries \eqref{eq:fc12} of the bistable domain are $F_{c1} \rightarrow 48$ and $F_{c2} \rightarrow 60$. Although the bistability persists in principle in this limit, figure~\ref{fig:bis} shows that the basin of attraction of the longitudinal orientation extends over all initial angles $0 \leq \ovl{\psi}_0<90^o$. As a consequence, the preferential orientation of a floater {with very small $b \ll 1$} is entirely governed by $F$, with no influence of the initial angle $\ovl \psi_0$, except for the singular initial condition $\ovl \psi_0 = 90^o$. This clearly appears by letting $b \rightarrow 0$ in equation \eqref{veff_psi}, yielding the simplified potential
\be
V(\ovl{\psi}) =  \frac{1}{4} \epsilon^2 \left ( - 1  +\frac{F}{{60} }   \right ) \cos^4 {\ovl \psi},
\label{eq:Vcos4}
\ee
and hence the simplified equation of motion \eqref{eq:psidd} mentioned in the Introduction. In the thin floater limit,  the longitudinal transverse transition occurs at $F_c = 60$ without bistability.

\section{Simplified model for very elongated floaters}

\label{sec:needle}

The asymptotic approach of the previous section is technical and hides much of the physics of the reorientation dynamics. In this section, we introduce a simplified model for strongly elongated floaters that leads to the same evolution equation \eqref{eq:psidd} and is more easily interpreted. 

\subsection{A simpler formula for the yaw moment}

We recall that both the rapid and slow motion of yaw angle $\psi$ are controlled by
\be
K_z =  - \int_{V_{sub}} (y-y_c) \, a_x \ d V,
\ee  
with $a_x = - \epsilon  e^z \cos (x-t)$ the horizontal acceleration of flow. We have seen at the end of the previous section that the limit of strongly elongated floaters $\delta_x \gg \delta_y \gg \delta_z $ is  adapted to capture the preferential orientation phenomenon, so let us exploit this information directly. If the floater is indeed thin (along $\wt{z}$) and not wide (along $\wt{y}$), we can ignore the  $\wt{y}$ and $\wt{z}$ variations in this moment integral and simplify it to
\be \label{Kzdef_needle_v0}
K_z \approx - \int_{-\delta_x/2}^{+\delta_x/2} 
(y-y_c) \,a_x  \, \delta_y  \underbrace{  \left(  \wt{\zeta}+\frac{\delta_z}{2} \right )}_{\wt{h} (\wt{x},t)}   d \wt{x} .
\ee 
For this formula to make sense, we 
must express $(y-y_c)a_x$ in terms of floater frame coordinates and  ignore the $\wt{y}$ and $\wt{z}$ variations. We also need to find the local submersion depth $\wt{h} (\wt{x},t) = \wt{\zeta}+({\delta_z}/{2}) $ along the long $\wt{x}$-axis  of the floater. From \eqref{mante_to_lab_lin} and ignoring all dependencies along $\wt{y}$ and $\wt{z}$, we find that  
\be \label{eq:simpl_tf_needle}
 x \approx x_c +  c_\psi \, \wt{x}  , \quad y \approx y_c +  s_\psi \, \wt{x}, \quad z \approx  z_c - \theta' \wt{x} 
\ee
along the long axis of the floater. The lever arm is then $y-y_c \approx  s_\psi \, \wt{x}$. The dependence of the field $a_x$ on $\wt{x}$ can be found by injecting this coordinate transform in the theoretical expression. Then, making use of a Taylor expansion we get
\ba
a_x &\approx & - \epsilon e^{z_c - \theta' \wt{x} } \cos ( x_c + c_\psi \wt{x}  - t )   \nonumber \\
& \approx& - \epsilon \left(1 + z_c - \theta' \wt{x} \right )  \left ( \cos ( x_c - t ) - c_\psi \wt{x}\,  \sin ( x_c - t ) \right )  + \ldots
\ea
up to $O(\epsilon^2)$. To find the local submersion depth $\wt{h}(\wt{x},t) $, we ignore the $\wt{y}$-dependencies in the general definition of $\wt{\zeta}$ in \eqref{htdef}. We also use the fact that $-\ovl{z}_{c} + (\delta_{z}/2) = \beta \delta_z = \ovl{h}$ is the equilibrium submersion depth. In this way, we find that the local submersion depth $\wt{h}(\wt{x},t) $ along the floater long axis is 
\be \label{wth_1}
\wt{h}(\wt{x}, t)   \approx   \ovl{h}  + (- z_c'  + \epsilon \sin (x_c-t ) )  + ( \theta'  +\epsilon c_\psi \cos (x_c-t ) \wt{x} - \frac{\epsilon}{2} c_\psi^2 \wt{x}^2    \sin (x_c-t) .
\ee
The wave surface at the scale of the floater is here approximated by a second order polynomial and this is sufficient in this simplified model. In this expression, we need to insert the first order motion $z_c'$ and $\theta'$. We already know the first order motion from the asymptotic theory, but let us show how we can alternatively find this motion using some simpler physical arguments. In the pressure force and moment, the buoyancy terms are the largest ones, much larger than the dynamical pressure terms. Hence, the floater is moving in such a way that it keeps its submerged volume nearly constant in time. This requires 
\be
  \int_{V_{sub}}   \, dV  \approx   \delta_y   \int_{-\delta_x/2}^{+\delta_x/2} 
  \wt{h} (\wt{x},t)   d \wt{x} \approx   m    \quad \Rightarrow \quad   \frac{1}{\delta_x}\int_{-\delta_x/2}^{+\delta_x/2} \wt{h}(\wt{x}, t)  d \wt{x} \approx \ovl{h} .
\ee
Evaluating the integral using the expression \eqref{wth_1}, we get the result
\be \label{eq:zcprimeneedle}
 z_c' \approx \epsilon \sin (x_c-t ) \left ( 1 - \frac{\delta_x^2}{24} c_\psi^2 \right ) .
\ee
This indeed corresponds to the expression of $z_c'$ that we have obtained from a more formal treatment of the equations of motion, in the limit $\delta_x \gg \delta_y$ (see equation~\eqref{zcprime}).  Interestingly, we also recover the shape-related $O(\epsilon \delta_x^2) $ correction that is quite crucial in the model.  To explain the angular motion $\theta'$, we can use a similar  argument. Due to buoyancy, the floater will rotate so as keep the Archimedes torque zero
\be \label{eq:thetaprimeneedle}
 \int_{V_{sub}} ( \bs{r} - \bs{r}_c) \times  \bs{e}_z   \,  d V  \approx \bs{0}\ \Rightarrow \    
\int_{-\delta_x/2}^{+\delta_x/2}  \wt{x}  \wt{h}(\wt{x}, t) d \wt{x}   \approx 0 \  \Rightarrow \  \theta'  \approx  - \epsilon c_\psi \cos (x_c-t )  .
\ee
This approximation is sufficient, the extra $O(\epsilon \delta_x^2)$ correction in $\theta'$ of equation \eqref{thetaprime} is not so important. Replacing these expressions of $z_c'$ and $\theta'$ into equation \eqref{wth_1}, we find the following approximation for the local shape of the waterline at the floater position
\be
\wt{h}(\wt{x}, t)   \approx   \ovl{h}  + \epsilon c_\psi^2   \left (\frac{\delta_x^2}{24} -\frac{\wt{x}^2}{2}    \right )    \sin (x_c-t) .
\ee
The waterline as seen from the floater center always takes the shape of a parabola, symmetrical around $\wt{x}=0$. 
Combining the elements together, we find a formula for the yaw moment $K_z$:
\ba
K_z & \approx&  - \int_{-\delta_x/2}^{+\delta_x/2} 
\underbrace{\wt{x} s_\psi }_{\text{local lever arm}} \, \underbrace{\left (- \epsilon \left(1 + z_c - \theta' \wt{x} \right )  \left ( \cos ( x_c - t ) - c_\psi \wt{x}\,  \sin ( x_c - t ) \right ) \right ) }_{ \text{local force density $f_x$}} \nonumber \\
& & \hspace*{2cm} \times \, \underbrace{ \left( \ovl{h}  + \epsilon c_\psi^2   \left (\frac{\delta_x^2}{24} -\frac{\wt{x}^2}{2}    \right )    \sin (x_c-t) \right ) }_{\text{local submersion $\wt{h}$}} \, \delta_y \,  d \wt{x} \label{Kz_needle}  .
\ea
This integral contains all the physics that explains the preferential orientation. The moment $K_z$ is the result of the local force density $f_x$ that varies along the floater's long axis applied on the local level arm. This force density is weighted by the local submersion depth $\wt{h}$ and this effect is thus more important as the floater is longer. 

\subsection{Evolution equation for $\ovl{\psi}$ }

We now evaluate the yaw moment \eqref{Kz_needle} up to $O(\epsilon^2)$ and write the evolution equation $\wt{I}_{zz} \ddot{\psi} = K_z$. Using $\wt{I}_{zz} \approx m \delta_x^2/12$ for our elongated floater, we find
\be \label{eq_psi_inter}
\ddot{\psi} \approx - \epsilon (1+ z_c) s_\psi c_\psi \sin (x_c-t) - \epsilon \theta' s_\psi \cos (x_c- t) + \frac{\epsilon^2 \delta_x^2}{30 \beta \delta_z} s_\psi c_\psi^3 \sin^2 (x_c -t ) .
\ee
We now inject in this equation the decomposition $x_c = \ovl{x}_c + x_c'$ and $z_c = \ovl{z}_c + z_c'$. In the first, $O(\epsilon)$ term of equation \eqref{eq_psi_inter}, we use a Taylor expansion:
\ba
&&   - \epsilon (1+ \ovl{z}_c + z_c') s_\psi c_\psi \sin (\ovl{x}_c + x_c'-t)   \\
&& \quad  =    - \epsilon (1+ \ovl{z}_c) s_\psi c_\psi \sin (\ovl{x}_c -t)  - \epsilon  z_c' s_\psi c_\psi \sin (\ovl{x}_c -t)  - \epsilon x_c'   s_\psi c_\psi \cos (\ovl{x}_c -t)  + O (\epsilon^3). \nonumber
\ea
As the floater is small and thin, we can approximate $\ovl{z}_c \approx 0$. The second and third terms of equation \eqref{eq_psi_inter} are already of order $O(\epsilon^2)$, so there we can use  $\sin^2 (x_c - t ) \approx \sin^2(\ovl{x}_c - t ) + O (\epsilon)$ and   $\cos (x_c - t ) \approx \cos (\ovl{x}_c - t ) + O (\epsilon)$. We then substitute the first order deviations 
\be
x_c' \approx \epsilon \cos (\ovl{x}_c -t ) \quad, \quad  z_c' \approx \epsilon \sin (\ovl{x}_c -t ) \quad, \quad \theta' = - \epsilon c_\psi \cos (\ovl{x}_c - t).
\ee
The expression of $x_c'$ can be found by integrating $\dot{x}'_c = u_x |_{z=0}$, which just means that, at lowest order, the floater translates as a fluid material particle on the surface. The quadratic correction of order $O(\epsilon \delta_x^2)$ is $z_c'$ is not needed here. After these reductions, we obtain
\be \label{eq:intermpsi}
\ddot{\psi} \approx - \epsilon  s_\psi c_\psi \sin (\ovl{x}_c-t) +  \epsilon^2 s_\psi c_\psi \left ( -1 + \frac{\delta_x^2}{30 \beta \delta_z}  c_\psi^2 \right  ) \sin^2 (\ovl{x}_c -t )
\ee
as evolution equation for $\psi$. We now inject the decomposition $\psi= \ovl{\psi} + \psi'$ and use a Taylor expansion to replace
\be
s_\psi c_\psi = \ovl{s}_\psi \ovl{c}_\psi  + \psi'  (\ovl{c}_\psi^2 -  \ovl{s}_\psi^2 ) + O (\epsilon^2),
\ee
yielding
\ba
\ddot{\ovl{\psi}} + \ddot{\psi}' &\approx&  - \epsilon  \ovl{s}_{\psi} \ovl{c}_{\psi} \sin (\ovl{x}_c-t) \nonumber \\
&&  - \epsilon  (\ovl{c}_{\psi}^2 -  \ovl{s}_\psi^2 ) \psi' \sin (\ovl{x}_c-t) + 
\epsilon^2 \ovl{s}_\psi \ovl{c}_\psi \left ( -1 + \frac{\delta_x^2}{30 \beta \delta_z}  \ovl{c}_\psi^2 \right  ) \sin^2 (\ovl{x}_c -t ).
\ea
At order $O(\epsilon)$, we identify the equation for the fast yaw angle excursion $\ddot{\psi}' =  - \epsilon  \ovl{s}_{\psi} \ovl{c}_{\psi} \sin (\ovl{x}_c-t)$, that has the solution 
\be
\psi' \approx \epsilon  \ovl{s}_\psi \ovl{c}_\psi \sin (\ovl{x}_c-t).
\ee
This expression is identical to that of the asymptotic model \eqref{psi_prime} in the limit $\delta_x \gg \delta_y$. Injecting this expression of $\psi'$ back into the equation and taking the average over the short time-scale, we find the second order evolution equation for $\ovl{\psi}$: 
\be
\ddot{\ovl{\psi}} \approx -\epsilon^2 \ovl{s}_\psi \ovl{c}_\psi^3 \bigg (  1  - \underbrace{\frac{\delta_x^2}{60 \beta \delta_z} }_{F/F_c} \bigg ).
\ee
The simplified model {reproduces exactly} the more formal asymptotic theory. 

\begin{figure}
\includegraphics[width=\textwidth]{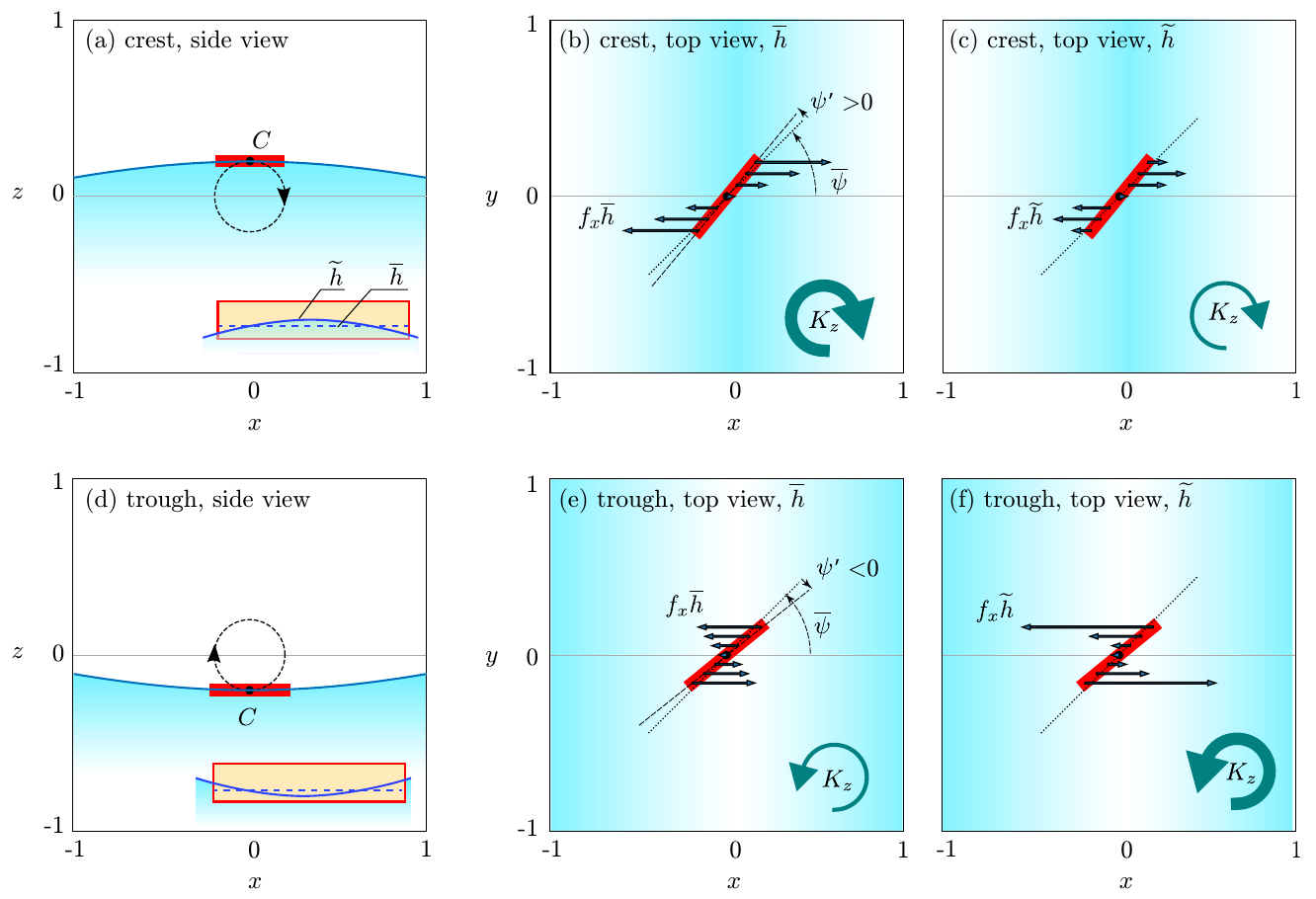}
\caption{Position, orientation and force distribution on the floater in wave crests and troughs. Here $\delta_x=0.5,\delta_z=0.01$, $\beta=0.2$, $\epsilon=0.2$ and $\ovl{x}_c =0$, $\ovl{\psi}=45^o$. (a,d) Side views on vertical $x-z$ plane. The circle shows the trajectory of the center of mass. The inset figures suggest the instantaneous, parabolic shape of the waterline near the floater.  (b,c,e,f) top views on $x-y$ plane. The arrows show the  instantaneous force distribution along the floater that creates the instantaneous moment $K_z$. In (b,e), without considering the variable submersion depth,  using $\overline{h}$. In (c,f) taking into account the variable submersion $\wt{h}$. 
\label{floater_physics_a}
}
\end{figure}

\subsection{Physical origin of preferential orientation}

The simplified model allows a better understanding of the physics that causes reorientation. Let us reconsider the moment $K_z$ in equation \eqref{Kz_needle}, expressed as the product of the local force density $f_x$ by the local level arm weighted by the local submersion depth $\wt{h}$. {Both quantities being rapidly oscillating in time, a nonzero product arises from a phase correlation between them, that we illustrate in figure~\ref{floater_physics_a}. Here, we show the floater at two wave phases, $t=\pi/2$ and $t=3\pi/2$ (wave crest and trough), and represent the force density $f_x$  as vector arrows; animations are available as Supplementary Materials.} The parameters are $\beta=0.2$, $\delta_x =0.5, \delta_z = 0.01$ and $\epsilon=0.2$, corresponding to a non-dimensional number $F = \delta_x^2 / \beta \delta_z = 125$, larger than the critical value $F_c=60$:  this floater will prefer the transverse position at late times. To illustrate the effect of the variation in submersion, we show in the second column [figures~ \ref{floater_physics_a}(b,e)], the
force distribution weighted by the {equilibrium, spatially uniform submersion depth $\overline{h}$}, whereas in the third column [figures~ \ref{floater_physics_a}(c,f)] it is weighted by the true, varying depth $\wt{h}$. 

We first consider the case where the depth variation is ignored [figures~\ref{floater_physics_a}(b,e)], an approximation acceptable for small floaters only. At the wave crest, the yaw angle is slightly larger than the mean value $\ovl{\psi}$ whereas at the trough, it is slightly smaller than $\ovl{\psi}$. The lever arm, $\wt{x} \sin \psi$, is therefore larger at the crest. We also see that the local force density $f_x$ is slightly larger at the crest than at the trough. Both effects impact the instantaneous moment in the same way. At the crest, the floater will experience  a negative (clockwise) moment $K_z <0$ that is  slightly larger than the positive (counter-clockwise) moment $K_z >0$ in the troughs, explaining why the part $-\wt{x} \sin \psi  f_x \ovl{h}$ in the integrand of $K_z$ is slowly pushing the floater towards the longitudinal position. 

Figures~\ref{floater_physics_a}(c,f) illustrates why including the varying submersion in the weighting of the force distribution changes this conclusion in the case of long floaters. We immediately see that there is clear influence of the variable submersion at the tips of the floater: the weighted force density $f_x \wt h$ is significantly changing in magnitude. As shown in figures~\ref{floater_physics_a}(a,d), at the wave crest, the extremities of the floater are less submerged, whereas in the troughs, they are more submerged. This locally changing submersion implies that the instantaneous moment is significantly decreased at the tips of the floater when they are at crest and increased at the tips when the floater is in a trough. The result for this floater with $F=125 > F_c$ is that the positive (counter-clockwise) moment $K_z>0$ acting on the floater at the troughs is significantly larger than the negative moment at the crests, resulting in a slow rotation towards the transverse position.

In the introduction, we have mentioned that the yaw angle motion of small floaters with $F<F_c$ is analogous to that of the Kapitza pendulum, a pendulum with an oscillating anchor point~\citep{kapitza_1951,landau,butikov_2001}. For such small floaters we have seen that the variation of the immersion depth can be ignored and in that case equation \eqref{eq:intermpsi} reduces to  
\be
F \ll F_c  \ : \ \ddot{\psi} \approx - \epsilon  s_\psi c_\psi \sin (\ovl{x}_c-t) - \epsilon^2 s_\psi c_\psi  \sin^2 (\ovl{x}_c -t ).
\ee
Using the fact that $s_\psi c_\psi  = (1/2) \sin 2 \psi$ and changing notation $2 \psi = \alpha$, we can  rewrite this as
\be
\ddot{\alpha} + \bigg  ( \underbrace{- \frac{\epsilon^2}{2}}_{\ovl{g}}   +  \underbrace{\epsilon \sin (\ovl{x}_c -t ) + O(\epsilon^2)}_{g'(t)}   \bigg ) \sin \alpha \approx 0 . 
\ee
This is identical to the equation of motion for the angle $\alpha$ of a Kapitza pendulum, written in the frame of reference attached to the anchor point. The Kapitza pendulum, analogue to our floater, would be exposed to a weak $O(\epsilon^2)$ downward external gravity $\ovl{g}$ and a larger $O(\epsilon)$ oscillatory acceleration $g'(t)$ that is due to the motion of the anchor point. Using the same multiple time-scale techniques as previously, we can find that this pendulum has two equilibria. The lower position $\ovl{\alpha} =0^o$ is stable and the top position, $\ovl{\alpha} = 180^o$ is unstable. Owing to the relation $\ovl{\psi} = \ovl{\alpha}/2$, this result is entirely equivalent to saying that short floaters prefer longitudinal positions, $\ovl{\psi} = 0^o$, and avoid transverse positions, $\ovl{\psi} = 90^o$.

\section{Mean yaw moment: comparison with the literature }
\label{sec:comparison}

In this section, we compare the second order yaw moment $\ovl{K}_z$ that we have calculated using our Froude-Krylov model to some existing results.  To allow comparison, let us start by writing the dimensional yaw moment according to our small floater, diffractionless theory. By multiplying the right hand side of equation \eqref{eq:psidd} with the dimensional moment of inertia $\beta \rho L_x^3 L_y {L_z}/ 12$ and $\omega^2 = gk $, we obtain
\be
\overline{K}_z = \frac{1}{12} \rho g  a^2 k^3 L_x^3 L_y  \left ( - \beta L_z + \frac{kL_x^2}{60 }\right ) \ovl{s}_\psi \ovl{c}_\psi^3 .
\label{eq:kzmd}
\ee
This formula only applies to parallelepiped floaters that are short with respect to the wavelength. As explained before, part of the moment (contribution $-\beta L_z$ in the parentheses) favors a longitudinal floater position (head-seas). This moment clearly depends on how deep the floater is submerged as the draft of our floater is $\ovl{h} = \beta L_z$. The other part of the moment (contribution $+ kL_x^2/60$ in the parentheses) favors a transverse floater position (beam-seas) and does not depend on the draft. Physically, this term relates to the spatial variation of the submersion along the long axis of the floater. 

\cite{newman_drift_1967} derived  an analytical formula [see his equation (55)] for the mean yaw moment on a slender parallelepiped. Written in our notations (wavenumber $K \rightarrow k$, width $B \rightarrow L_y$, length $L \rightarrow L_x$, angle of incidence  $\beta =  - \ovl{\psi }$), this mean yaw moment is 
\be
\overline{K}_z^{\text{\, Newman}} = \frac{1}{2} \rho g k a^2 L_x^2 L_y \sin \ovl{\psi} \, j_1 \left( \frac{1}{2} k L_x  \ovl{c}_\psi \right ) j_2 \left( \frac{1}{2} k L_x  \ovl{c}_\psi \right ),
\label{eq:kzmn}
\ee
with $j_1 $ and $j_2$ the spherical Bessel functions that relate to the \text{sinc} function and its derivatives. These functions oscillate with approximate period of $2 \pi$ and they decay for large argument. Hence, as we increase $kL_x$, this mean yaw moment can change multiple times in sign to gradually vanish in the limit of $k L_x \rightarrow \infty$. In the limit of small $kL_x $  where our theory applies, we can replace the spherical Bessel functions with their small argument asymptotic expansions. From \cite{abramowitz1948handbook}, equation (10.1.2), we have  $j_1 (x) j_2 (x) \approx x^3 / 45$ for small $x$ (rather than $x^3/3$ as written in \cite{newman_drift_1967}) and the mean yaw moment \eqref{eq:kzmn} reduces to
\be
k L_x \cos \ovl{\psi} \ll 1 ~ : \quad \overline{K}_z^{\text{\, Newman}} \approx \frac{1}{12} \rho g  a^2 k^3 L_x^3 L_y \left (  \frac{kL_x^2}{60 }\right )\ovl{s}_\psi \ovl{c}_\psi^3 .
\ee
Interestingly, we find that Newman's mean yaw moment formula contains exactly one term of our formula (\ref{eq:kzmd}), the one that favors the transverse position and relates to the spatial variation of submersion. The other term, related to the draft $\ovl{h}=\beta L_z$, is absent. This missing term in Newman's formula explains why he did not predict a stable longitudinal position for short floaters. 

The fact that our theory based on the Froude-Krylov assumption recovers the short floater limit of Newman's theory suggests that the diffracted/radiated wave correction does not contribute at leading order to the mean yaw moment on small floaters. This is not directly visible in a far field theory like that of Newman, because the existence of a diffracted/radiated wave is essential in that method: without radiated wave no angular moment is carried away. We found it intruiguing that a diffractionless model (ours) can produce the same result as a theory (Newman's) where diffraction is essential. This observation motivated us to investigate what a diffractionless approach produces as mean yaw moment, for longer floaters, with $\delta_x > 1$.  In the Appendix \ref{sec:appendix}, we extend the model of section \ref{sec:needle} to the case of long floaters. Ignoring diffraction with long floaters is risky, but nevertheless an interesting exercise as it turns out that one can calculate Newman's mean yaw moment formula \eqref{eq:kzmn} from a purely diffractionless, Froude-Krylov approach. This suggests that diffractive corrections of the wave are as absent in Newman's mean yaw moment formula as they are absent in our Froude-Krylov theory. Explaining this result requires a deeper investigation that is beyond this article. It might be that  the use of a slender body approximation filters the diffractive corrections in the mean yaw moment formula of Newman.

\begin{figure}
\includegraphics[width=\textwidth]{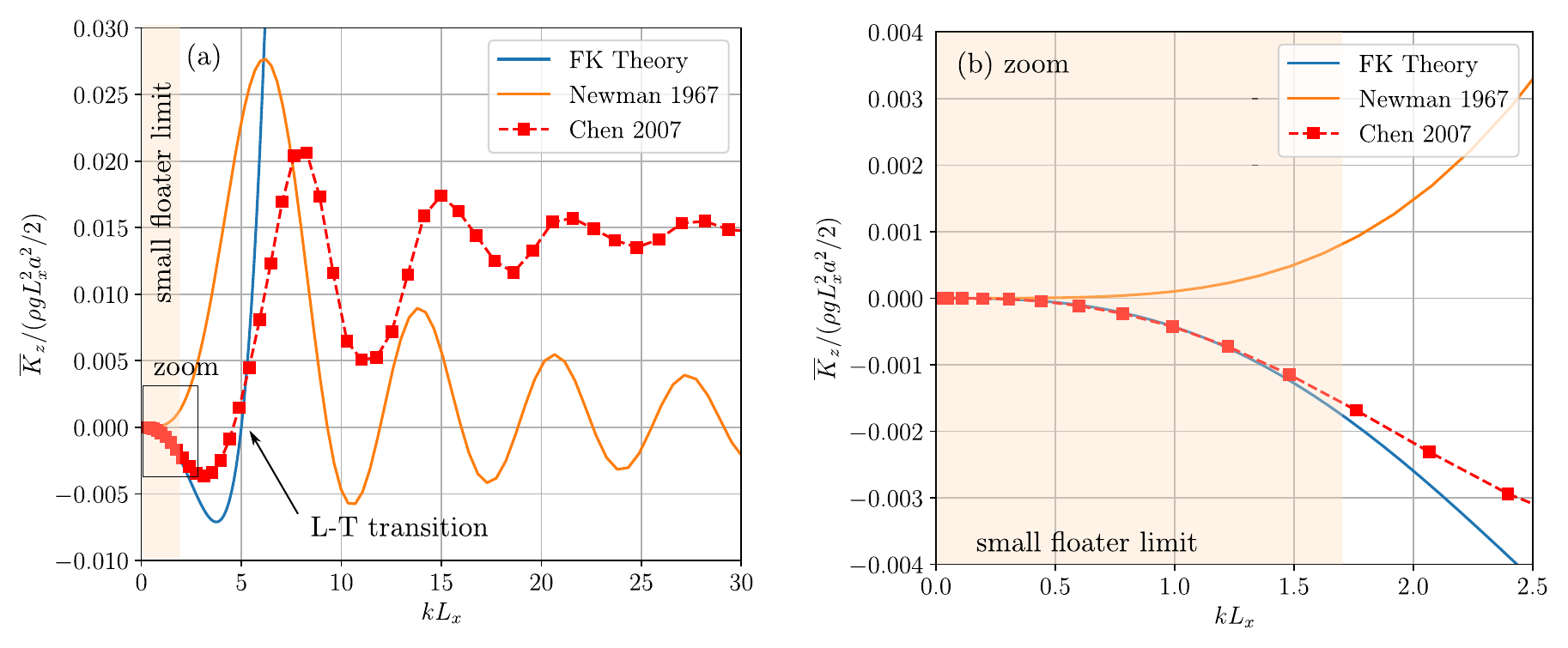}
\caption{(a) Dimensionless mean yaw moment as a function of $kL_x$, and (b) zoom at small $kL_x$. Red squares: Data from the boundary element simulations of \cite{chen2007middle}, computed for a 300-m long floating structure. Blue line: present Froude-Krylov theory \eqref{eq:kzmd}, valid in the limit $kL_x \leq 1$, when diffraction and radiation is negligible. The longitudinal-transverse transition, where $\ovl{K}_z=0$ (or $F=F_c$), is approximately the same for both methods.
\label{chen_compar} }
\end{figure}

In his {\it middle-field} formulation, \cite{chen2007middle} introduced a numerical approach to calculate first order motion and  second order load on quite general floating structures. The potential field around the floating body is calculated using the boundary element method and includes diffracted and radiated waves.  In one numerical application, Chen considers a very large floating platform, called FPSO (Floating Production Storage Offloading)-unit, that is close to a parallelepiped with dimensions  $L_x=300$~m by $L_y =50$~m and submerged over $\ovl{h} = \beta L_z = 25$ m  -- huge compared to our centimeter scale parallelepiped floaters --, placed in waves with angle of incidence $\ovl{\psi} = -165^o$. In common sea conditions,  wavelengths are usually shorter than $300$~m, so  we expect diffraction and wave radiation corrections, ignored in our theory, to be important.  In figure~\ref{chen_compar}, we compare the mean yaw moment calculated by Chen's BEM approach (red squares) to our small floater Froude-Krylov theory \eqref{eq:kzmd} (blue line) and to Newman's (1967) formula \eqref{eq:kzmn} (orange line). As $kL_x$ increases, the mean yaw moment calculated by Chen is first negative, then changes sign and oscillates at larger $kL_x$ to saturate at a constant value in the short wavelength limit $kL_x \rightarrow +\infty$. This saturation to a non-zero value is due to the non-symmetrical shape of the FPSO. Newman's formula captures the correct order of magnitude but differs everywhere from Chen's calculation. A comparison to our small floater theory for rectangular parallelepipeds only makes sense in the $kL_x <1$ limit, here emphasized by the zoom in figure~\ref{chen_compar}(b). There we see that the $- (kL_x)^4$ trend at small $k L_x$ is very well reproduced by our theory, a trend that is absent in Newman's formula. At larger $kL_x$, there is a strong departure of our formula from the simulation. 


\section{Conclusion}

In this paper we have studied the preferential orientation of elongated floaters in propagating gravity waves, focusing on the case of small floaters that have $kL_x < 1$. Experiments in this regime indicate that short and deeply immersed floaters align longitudinally, along with the direction of propagation, whereas long and weakly immersed floaters prefer to align transversely, along with the wave crests and troughs.

We have shown that this preferential orientation can be modeled using a strongly idealised diffractionless, Froude-Krylov approach, that ignores finite depth effects, viscous effects, capillary effects and steady streaming flows. Numerically integrating the {resulting} equations of motion of this model, we have found preferential orientations that compare well to experimental observations. We then went on with this model and derived an asymptotic description of motion, in the limit of small wave slope $\epsilon \ll 1$ and for small floater size $\delta \ll 1$. For strongly elongated floaters with height, width and length ordered as $\delta_z \ll \delta_y \ll \delta_x \ll 1$, we could compute the second order mean yaw moment {that lead us to equation} \eqref{eq:deff} for the slow motion of the yaw angle. Although this idealised, dissipation-less equation is not sufficient to model realistic yaw angle motion, it {shows} that the {preferential orientation} of small elongated floaters is nearly independent of $\epsilon$ and mainly controlled by the non-dimensional number $F = k L_x^2 / \beta L_z $. When $F< F_c$, the floater will favor a longitudinal orientation, when $F> F_c$, the floater will favor a transverse orientation. In the theory, we have found that critical number $F_c = 60$, while experiments give $F_c =50 \pm 15$ or even lower $F_c \simeq 35 \pm 10$ if we use the experimentally measured submersion depth (modified by capillarity).   

Considering the simplifying assumptions of the theory, there are many effects that can contribute to the slight difference between experiments and theory. First, the wave in the experiments is not a {perfect} propagating wave {in deep water, but has finite depth effects and some wave reflection  due to imperfect attenuation at the end of the channel. Second, viscous forces are certainly acting on our small floater and they may alter the mean yaw moment. Including viscosity can be done but requires a modeling of the Stokes boundary layer under the floater. Third, we have seen that surface tension modifies the equilibrium immersion depth, but it can also introduce additional horizontal forces when the floater is moved by the wave. Fourth, we did not include the effect of a steady streaming flow. Fifth, we did not include diffraction, {which is questionable for $k L_x$ approaching $O(1)$}. Sixth and finally, the prediction $F_c = 60$ holds only in the limit of very elongated floaters, $\delta_y / \delta_x \ll 1$. Including finite width effects slightly lowers the transition value, and introduces a bistability in the equilibrium positions in the vicinity of the transition, that may contribute to the experimental spread.  

To gain deeper insight into the physics of the floater orientation, we have shown that the evolution equation  \eqref{eq:deff} for the slow motion of the yaw angle can also be found using a simpler approach. A careful inspection of equation \eqref{Kz_needle} and the subsequent analysis allows to separate the different contributions to the mean yaw moment and reveals a physical meaning of the number $F$,
\be
F = \frac{\text{Mean yaw moment due to spatially varying submersion $h' = \wt{h} - \ovl{h}$}}{\text{Mean yaw moment due to first order displacement $x_c',\theta',z_c' $}}.
\ee
This {is supported by the examination of the instantaneous force distribution along the floater when the effect of the variable submersion is included or discarded in the computation of the moment, as sketched in figure \ref{floater_physics_a}.} Short floaters see little variation in submersion depth along their long axis and experience a mean moment that favors the longitudinal position.  This mean moment arises from a phase correlation between the oscillating buoyancy force and the oscillating lever arm, a feature shared with the classical Kapitza pendulum. For longer floaters, the variation of the submersion along the floater has a strong effect on the instantaneous moment, that is significantly decreased in crest positions (the tips are less submerged) and  increased in trough positions (the tips are more submerged). Since in the trough position, the instantaneous moment always pushes towards the transverse position, long floaters prefer to take transverse positions. 

We finally compared our mean yaw moment formula for small floaters to previously published results. Compared to Newman's theory, we have identified an additional contribution to the mean yaw moment that varies linearly with the draft $\overline{h}$. The longitudinal-transverse transition for short floaters is due to this extra contribution. Newman's prediction, that slender structures are stable {in transverse orientation (``beam-seas'')}, does not apply to floaters that are small with respect to the wavelength. Comparing our theory to  Chen's boundary element calculations, we obtain excellent agreement in the {limit of floater length much smaller than the wavelength}. Away from this limit, diffraction and radiation are no longer negligible and our simplified theory breaks down. 

The present study on the mean yaw moment can be continued in several directions. Extending the model of section \ref{sec:needle} to elongated floaters with non-symmetrically distributed mass and a realistic hull is not  difficult. It would be interesting to compare such a model to results obtained using the methods of \cite{chen2007middle}. Another interesting perspective relates to our appendix: what is the precise contribution of diffractive correction to the mean yaw moment? Finally, we can also draw some parallels between our work and recent studies in relation to the problem of plastic waste transport by waves. The effect of shape on the mean motion of non-spherical objects in wave flows has been investigated in several studies, but limited to fully submerged, neutrally buoyant ellipsoids~\citep{dibenedetto_transport_2018,dibenedetto_preferential_2018,dibenedetto_orientation_2019}. A similar preferential orientation phenomenon is observed there too and it would be interesting to study whether the physical origin of the orientation is  different or similar. 


\section*{Supplementary data}

Supplementary material and movies are available at ...

\section*{Acknowledgments}

We are grateful to  J. Andriamampianina, S. Courrech du Pont, A. Eddi, M. Le Boulluec,  L. Martin-Witkowski and  M. Rabaud for fruitful discussions, and X. Chen for sharing his numerical data on the mean yaw moment. We thank  A. Aubertin, L. Auffray, J. Amarni, R. Pidoux and J. Zhang for experimental help.

\section*{Declaration of Interests} 

The authors report no conflict of interest.

\begin{appendix}

\section{Calculating Newman's mean yaw moment from a diffractionless model}

\label{sec:appendix}

Newman's original formula for the mean yaw moment on slender floaters of arbitrary length was given in equation \eqref{eq:kzmn}. This formula is derived from a global angular momentum balance and expresses the mean yaw moment on the floater as a function of the angular momentum that radiates away from the floater, in the far field. Using Green function theory and a slender body approximation, Newman achieved to calculate the  Kochin function that appears in the mean yaw moment formula. This far-field method essentially relies on the existence of a diffracted/radiated wave and certainly suggests that diffraction is taken into account. As we show here this may well be different, Newman's formula \eqref{eq:kzmn} can be derived from a Froude-Krylov model that ignores diffraction. 

To demonstrate this strange result, we return to the simplified theory of section \ref{sec:needle}, where we calculated the yaw moment on an elongated floater using the integral
\be
K_z \approx - \int_{-\delta_x/2}^{+\delta_x/2} 
(y-y_c) \,a_x  \, \delta_y  \, \wt{h} (\wt{x},t)   d \wt{x} .
\ee 
In writing this formula, we ignore diffraction and we also suppose that the floater is very thin and not wide, $\delta_z \ll \delta_y \ll 1$, to replace the integration over $\wt{y}$ and $\wt{z}$ with the factor $\delta_y  \, \wt{h} (\wt{x},t) $. In section \ref{sec:needle}, we have used Taylor expansions to replace the local fluid acceleration $a_x$ and local submersion depth $\wt{h} (\wt{x},t)$ with polynomials. It turns out that we can also calculate all these integrals exactly, without making use of these polynomial approximations. Hence we can explore what our theory suggests in the case of
\be
\delta_z \ll \delta_y \ll 1, \qquad  \delta_x   \text{  arbitrary} .
\ee
Ignoring diffraction is no longer advisable when the floater gets long, $\delta_x > 1$, as in figure~\ref{fig:long}. Nevertheless, we have found it interesting to test {what mean yaw moment a diffractionless approach would produce}  in this case of longer floaters. The full calculation is quite complex but there is one result that is really worth mentioning here. Newman's mean yaw moment formula \eqref{eq:kzmn} can be found as the part of the mean yaw moment that relates to the spatial variation of submersion. More precisely, it is exactly equal to 
\be \label{eq:kznewman_new}
\ovl{K}_z^{\, \text{Newman,nd}} \approx \overline{- \int_{-\delta_x/2}^{+\delta_x/2} 
(y-y_c) \,a_x  \, \delta_y  \, {h}' (\wt{x},t)   d \wt{x} }
\ee 
in non-dimensional form. Here we denote $h'  (\wt{x},t)  = \wt{h} (\wt{x},t)  - \ovl{h} = O (\epsilon)$ the local deviation from the equilibrium submersion. We explain the essential steps in the calculation that lead to this result. We first use the simplified transform of equation~\eqref{eq:simpl_tf_needle} and the definition $a_x = - \epsilon e^z \cos (x-t)$ to replace
\be  \label{eq:axnewneedle}
-(y-y_c)  a_x \approx s_\psi \wt{x} \, \epsilon \underbrace{e^{z_c - \theta' \wt{x}}}_{\approx 1 } \cos (x_c -t + c_\psi \wt{x}) 
\ee
in the integrand. The exponential factor can here be simplified to $1$ because we only need an $O(\epsilon)$  to evaluate \eqref{eq:kznewman_new}, considering that $h' = O (\epsilon)$.  
\begin{figure}
\begin{center}
\includegraphics[width=0.6\textwidth]{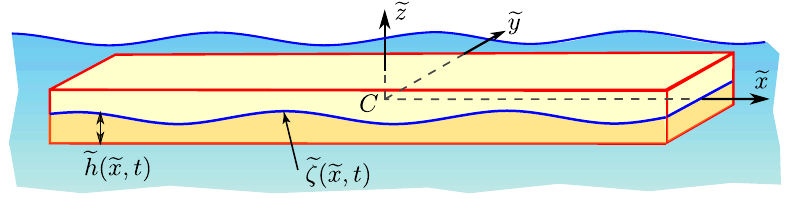}
\end{center}
\caption{Long floater in a wave.
\label{fig:long} }
\end{figure}
To find $h'(\wt{x},t)$, we need to calculate the local submersion $\wt{h}(\wt{x},t)$ along the axis of the long floater and this is done using the same physical arguments as in section \ref{sec:needle}. We inject $x \approx x_c + {c}_\psi \wt{x}$ and $z \approx z_c - \theta' \wt{x} + \wt{z}$ in the definition of the surface, $ z= \epsilon \sin (x-t) $, and rewrite this as $\wt{z}= \wt{\zeta} (\wt{x},t)$. This yields $ \wt{\zeta} (\wt{x} ,t) \approx - z_c + \theta' \wt{x} +  \epsilon \sin (x_c - t + c_\psi \wt{x}) $. The local submersion depth is by definition $\wt{h}  = \wt{\zeta} + \delta_z /2 $, and replacing $z_c = \ovl{z}_c + z_c'$ and  $\ovl{h} = - \ovl{z}_c  + \delta_z/2  $, we find 
\be
\wt{h}(\wt{x}, t)   \approx   \ovl{h} - z_c' + \theta' \wt{x} +  \epsilon \sin (x_c - t + c_\psi \wt{x}).
\ee
A Taylor expansion for small $\wt{x}$ gives \eqref{wth_1}. In this equation, we still need to determine $z_c'$ and $\theta'$ and this is done in the same way as before. By {imposing} that the submerged volume remains constant in time, we can fix $z_c'$, 
\be
 \delta_y   \int_{-\delta_x/2}^{+\delta_x/2} 
  \wt{h} (\wt{x},t)   d \wt{x} \approx   m    \quad \Rightarrow \quad
z_c' = \epsilon \, \text{sinc} \left ( \frac{c_\psi \delta_x }{2}  \right ) \sin (x_c - t) .
\ee
Here $\text{sinc} ( \alpha) = (\sin \alpha )/ \alpha$ is the sinc function, that also relates to the spherical Bessel function  $j_0 (\alpha) = \text{sinc} ( \alpha)$. A small  argument expansion of this sinc function in $z_c'$ indeed yields equation~\eqref{eq:zcprimeneedle}. Due to this sinc function,  the vertical motion $z_c'$ decreases in magnitude in an oscillating manner as $\delta_x$ becomes large. To find $\theta'$, we express that the instantaneous Archimedes moment vanishes: 
\be
\int_{-\delta_x/2}^{+\delta_x/2}  \wt{x}  \wt{h}(\wt{x}, t) d \wt{x}   \approx 0 \  \Rightarrow \  \theta'  \approx \frac{6 \epsilon}{\delta_x} \text{sinc}' \left ( \frac{c_\psi \delta_x }{2}  \right ) \cos (x_c -t ),
\ee
with $\text{sinc}' ( \alpha) = ( \alpha \cos \alpha - \sin \alpha)/\alpha^2  $ the derivative of the \text{sinc} function. In the limit of small argument, we have $\text{sinc}' ( \alpha) \approx - \alpha/3$ and this gives equation \eqref{eq:thetaprimeneedle} for the small floaters. With these expressions of $z_c'$ and $\theta'$ we find the local submersion depth $\wt{h}$ or more precisely, the deviation $h' = \wt{h} - \ovl{h}$ as  
\ba 
{h}'(\wt{x}, t)    &\approx &
\epsilon \sin (x_c - t) \left [ \cos (c_\psi \wt{x}) - \text{sinc} \left ( \frac{c_\psi \delta_x }{2}  \right ) \right ] \nonumber \\ 
&& \quad  + \epsilon \cos (x_c - t) \left [ \sin (c_\psi \wt{x}) + \frac{6 \epsilon}{\delta_x} \text{sinc}' \left ( \frac{c_\psi \delta_x }{2}  \right ) \right ]. \label{eq:hprime}
\ea
With equations \eqref{eq:axnewneedle} and \eqref{eq:hprime}, we have all the necessary to calculate the integral of equation \eqref{eq:kznewman_new},
\be
\ovl{K}_z^{\, \text{Newman,nd}} \approx - \frac{\epsilon^2 \delta_x^2 \delta_y}{4} s_\psi  \,  \text{sinc}' \left ( \frac{c_\psi \delta_x }{2}  \right )   \left [ \text{sinc}   \left ( \frac{c_\psi \delta_x }{2}  \right )  + 3 \, \text{sinc}''  \left ( \frac{c_\psi \delta_x }{2}  \right ) \right ]  .
\ee
We may replace $\psi \approx \ovl{\psi}$ in this formula as we only need the $O(\epsilon^2)$ moment. Using the spherical Bessel function properties
\be
\text{sinc} ( \alpha) = j_0 (\alpha) , \quad   \text{sinc}' ( \alpha) = - j_1 (\alpha) , \quad   \text{sinc} ( \alpha) + 3 \text{sinc}'' ( \alpha) = 2 j_2 (\alpha),
\ee
we can rewrite this mean yaw moment as 
\be
\ovl{K}_z^{\, \text{Newman,nd}} \approx  \frac{\epsilon^2 \delta_x^2 \delta_y}{2} \ovl{s}_\psi  \,  j_1 \left ( \frac{\ovl{c}_\psi \delta_x }{2}  \right )  j_2 \left ( \frac{\ovl{c}_\psi \delta_x }{2}  \right )  .
\ee
and if we multiply this with the dimensional factor $\rho g / k^4$, we exactly get Newman's formula \eqref{eq:kzmn} for the mean yaw moment on an elongated parallelepiped floater. 

\end{appendix}

\bibliographystyle{jfm}
\bibliography{ref_floater}

\end{document}